\documentclass{aa}  

\usepackage{graphicx}
\usepackage{txfonts}
\usepackage{hyperref}
\begin{document} 
\defcitealias{DK11}{DK11}
\defcitealias{Hee+15}{H+16}

   \title{Sedna and the cloud of comets surrounding the solar system\\in Milgromian dynamics}


   \author{R. Pau\v{c}o
          \and
          J. Kla\v{c}ka
          }

   \institute{Faculty of Mathematics, Physics and Informatics, Comenius University in Bratislava,
    Mlynsk\'{a} dolina, 811 02 Bratislava\\
              \email{pauco@fmph.uniba.sk}
             }


 
\abstract
{We reconsider the hypothesis of a vast cometary reservoir surrounding the Solar System
- the Oort cloud of comets - within the framework of Milgromian Dynamics (MD or MOND). 
For this purpose we built a numerical model of the cloud assuming QUMOND, a modified gravity theory of MD. 
In the modified gravity versions of MD, the internal dynamics of a system is influenced by the external gravitational field in which the system is embedded, even when this external field is constant and uniform, a phenomenon dubbed the external field effect (EFE).
Adopting the popular pair $\nu(x)=[1-\exp(-x^{1/2})]^{-1}$ for the MD interpolating function and $a_{0}=1.2\times10^{-10}$ m s$^{-2}$ for the MD acceleration scale, we found that the observationally inferred Milgromian cloud of comets is much more radially compact than its Newtonian counterpart. 
The comets of the Milgromian cloud stay away from the zone where the Galactic tide can torque their orbits significantly. However, this does not need to be an obstacle for the injection of the comets into the inner solar system as the EFE can induce significant change in perihelion distance during one revolution of a comet around the Sun. 
Adopting constraints on different interpolating function families and a revised value of $a_{0}$ (provided recently by the Cassini spacecraft), the aforementioned qualitative results no longer hold, and, in conclusion, the Milgromian cloud is very similar to the Newtonian in its overall size, binding energies of comets and hence the operation of the Jupiter-Saturn barrier. However, EFE torquing of perihelia still play a significant role in the inner parts of the cloud. Consequently Sedna-like orbits and orbits of large semi-major axis Centaurs are easily comprehensible in MD. In MD, they both belong to the same population, just in different modes of their evolution. 
}

\keywords{comets: general --- Galaxy: general, solar neighborhood --- gravitation --- Oort Cloud}

\maketitle
%

\section{Introduction}

Our present day theoretical framework of the Universe is the general theory of relativity (GTR; with a final piece in \cite{Ein15}), celebrating 100 years of its existence. GTR can, at an appropriate limit, be well substituted with Newtonian gravity since it was constructed for this, thus at some point GTR was adjusted to observations made in Newton's era. To explain the modern large-scale observations of the Universe with GTR, we have to insist on a nearly flat non-monotonously accelerating Universe filled with never directly observed ingredients, the so-called dark energy (well represented by the cosmological constant $\Lambda$) and non-baryonic dark matter (DM, or CDM for cold dark matter), both having very finely-tuned properties (e.g. \citealp{Cop+06, FM13}).

Unfortunately the $\Lambda$CDM model of the Universe is mute in addressing observed dynamical regularities of galaxies, the building blocks of the Universe: the baryonic Tully-Fisher relation \citep{TF77, McG+00, McG05b}, the Faber-Jackson relation \citep{FJ76, San10}, or the mass discrepancy-acceleration correlation \citep{McG04, McG05a}. These observations reveal a strong coupling between the baryonic matter and the hypothetical DM. Moreover, they self-consistently point to the existence of a special acceleration scale (\citealt{FM12}). 

Observations of our closest cosmic neighbourhood, the Local Group, highly disfavour the standard cosmology based on the particle dark matter (e.g. \citealp{Kro+10, Kro12}). One of the observations that is hard to accommodate within $\Lambda$CDM, even after baryonic physics is incorporated into the model, is the highly anisotropic distribution of the Local Group members - existence of thin co-orbiting planes of satellites around the Galaxy and M31 \citep{Paw+12b, Paw+13, Paw+14, Paw+15, Iba+13}. It has recently been discovered that similarly anisotropic distributions of satellites are possibly common in a low redshift Universe ($z < 0.05$; \citealp{Iba+14, Iba+15}).
All these issues signal that, after 100 years, we have probably reached the boundaries of GTR and it happened very naturally with empirical progress. Thus we should try to find and test a new theory that provides a better explanation for present-day observations.

The aforementioned galactic phenomenology can be well explained within the framework of
Milgromian dynamics (MD or MOND; \citealp{Mil83b, FM12} for a review of 30 years of its evolution). For instance, the thin co-orbiting planes of Local Group satellites can be a by-product of a past close fly-by that the Galaxy and M31 have undergone about 7 - 11 Gyr ago \citep{Zha+13, Paw+12a}. Thus, we can make the claim that the new theoretical framework of the Universe that we are looking for will explain why everything happens as if galaxies are Milgromian, and not Newtonian objects. 

The current status of MD is quite analogous to Newton's gravitational law, explaining the Kepler laws of planetary motion, in Newton's era: MD has strong predictive power although its parent (generally-covariant) theory is still absent (\citealt{FM12}). MD proposes a modification of dynamics that is most apparent in low-acceleration regions of astrophysical systems. In MD, a test particle in a point mass gravitational field accelerates towards the point mass with magnitude $(g^{N}a_{0})^{1/2}$ if $g^{N}\ll a_{0}$, where $g^{N}$ is expected Newtonian gravitational acceleration and $a_{0}$ is a constant with units of acceleration. The constant $a_{0}\sim10^{-10}$ m s$^{-2}$ plays the role of a moderator and vice-versa when $g^{N}\gg a_{0}$ the classical limit is recovered. However, MD also states (at least when considered as modified gravity) that the internal gravitational dynamics of a system is influenced by the existence of a constant external gravitational field\footnote{In spatially varying gravitational field we also have standard tidal effects.} in which the system is embedded \citep{Mil83b}. In MD, external gravity does not decouple from internal dynamics as it does in Newtonian dynamics; the strong equivalence principle is apparently broken. This so-called external field effect (EFE) can attenuate or erase MD effects in the presence of an external field of magnitude that is larger than $a_{0}$, even when internal accelerations are well below $a_{0}$, see Sect. \ref{Sec:EFE}.

Many of new comets entering the inner solar system can be good probes of modified dynamics as we expect them to originate at large heliocentric distances where the Sun-comet acceleration is very small\footnote{But note that EFE always attenuates classical MD effects, see Sect. \ref{Sec:EFE} for discussion.}. 
When astronomers observing motion of new comets entering the inner solar system interpret these observations in the framework of Newtonian dynamics (Newtonian astronomers) they end up with the idea of a vast reservoir (radius $\sim$ 100 kau) of bodies, the Oort cloud (OC; \citealp{Opi32, Oor50}), from which the comets are steadily replenished. In the language of the Newtonian orbital elements this happens because they find: (1) a sharp peak in the distribution of the original (i.e. before entering the planetary zone) reciprocal semi-major axes $0<1/a_{orig}\lesssim10^{-4}$ (i.e. orbital energies), and, (2) nearly isotropically distributed perihelia directions.
We reserve the terms ``near-parabolic comet'' and ``Oort-spike comet'' for a comet with semi-major axis greater than 10 kau and perihelion distance between 0 and $\sim8$ au (i.e. to be observable), as derived by a Newtonian astronomer.

In this paper, we investigate the change of the view about the solar system cometary reservoir when Newtonian dynamics is substituted with Milgromian.
We consider the exclusively quasi-linear formulation of MD (QUMOND; \citealt{Mil10}), the classical modified gravity theory that was constructed in the spirit of MOND (\citealt{Mil83b}).
We emphasize that the comet is observable only in the deep Newtonian regime where gravity is much larger than the MD threshold value $a_{0}$. 
The basic structure of the hypothetical cloud in MD can be thus probed by tracing the motion of Oort-spike comets back in time, with the actual observations (positions and velocities) serving as the initial conditions. Extending our mainly qualitative analysis into quantitative type presents a profound test of MD.

In the rest of Sect. 1 we briefly review the classical picture of the cometary reservoir. In Sect. \ref{MD} we introduce a quasi-linear formulation of MD (QUMOND) and the numerical procedure of ``how to move things'' in QUMOND. Sect. \ref{models} presents various models of the solar system that is nested in the local Galactic environment, as considered in this paper. The crude picture of the Milgromian OC (MOC) is presented in Sect. \ref{MilgromianOC}. In Sect. \ref{simul} we examine past QUMOND trajectories of 31 observed near-parabolic comets. In Sect. \ref{XXZ} we investigate torquing of perihelia induced by the MD's EFE. Constraints on the MD interpolating function families, as recently found by \citet{Hee+15}, are taken into account in Sect. \ref{if}. We conclude and discuss our results in Sect. \ref{sum}.

\subsection{The classical Oort cloud}\label{classicalOC}
We refer to the OC, whose existence, size and structure are inferred by a Newtonian astronomer as ``the classical OC''.

The standard picture is that the OC with a radius of several tens of kau is a natural product of an interplay between the scattering of planetesimals by the giant planets - inflating bodies' semi-major axes - and tidal torquing by the Galaxy, and random passing stars - lifting bodies' perihelia out of the planetary zone \citep{Dun+87, Don+04}. 
Vice-versa reinjection of these bodies
into the inner solar system is moderated by the same dynamical agents \citep{HT86, KQ09}. The pivotal role of the Galactic tide, in both enriching and eroding the OC, was fully recognized after the paper of \citet{HT86}. Their simplified analytical theory of the Galactic disk tide, taking only its vertical component into consideration (if we assume that the Galactic equatorial plane is ``horizontal''), as the radial components are nearly an order of magnitude weaker, reveals that the effect of the tides is analogous to the effect of the planets on comets of shorter periods -- causing the Lydov-Kozai cycles. The vertical component of the comet's orbital angular momentum is conserved and comets follow closed trajectories in the $q-\omega$ plane ($q$ is the perihelion distance and $\omega$ is the argument of perihelion). Thus, $q$ can be traded for a Galactic inclination back and forth, while $\omega$ librates around some fixed value. Since the component of the tidal force that brings comets into visibility is $\sim\sin(2b_{G})$, where $b_{G}$ is the galactic latitude of comet's aphelion, a comet experiences the most rapid changes of $q$ per orbit when $b_{G}=\pm\pi/4$, while when $b_{G}=0$, or $b_{G}=\pm \pi/2$, the changes in the perihelion distance are nil \citep{Tor86}.
Using a sample of long periodic comets (LPCs), with periods longer than 10 000 yrs and accurately known original orbits, \citet{Del87} also noted these features observationally in the distribution of $b_{G}$ among the sample comets, confirming the significance of the Galactic tide.

The comets with $q<15$ au are usually considered lost from the OC, either to the interstellar region or a more tightly bound orbit, owing to planetary perturbations (phenomenon also called Jupiter-Saturn barrier). The planetary kick they receive is typically much larger than the width of the Oort spike. Thus, to be observable, a comet has to decrease its perihelia by at least $\sim$10 au during the revolution that precedes its possible discovery from the zone where planets have a minor effect down to the observability zone (typically less than 5 au from the Sun). Only comets with $a>20 - 30$ kau (defining outer OC; $a$ is the semi-major axis) experience large enough tidal torque to cause this kind of large decrease in $q$ in one revolution (e.g. \citealp{Don+04, Ric14}). But, there are many observed Oort spike comets with much smaller semi-major axes (\citealp{DK11}, hereafter \citetalias{DK11}).
The concept of the Jupiter-Saturn barrier should actually be revised as about 15$\%$ of the near-parabolic comets can migrate through it without any significant orbital change (\citetalias{DK11}; \citealp{DK15}).

\citet{KQ09} demonstrate the importance of a special dynamical pathway capable of delivering inner OC bodies (initial $a<20$ kau, often even $<10$ kau) into the observable orbits -- but at first into the outer OC region $a>20$ kau -- by a cooperation between the planetary perturbations and the Galactic tide. According to \citet{KQ09}, the new comets entering the inner solar system could originate in both the inner, and the outer, OC, with nearly equal probability.
 
Passing Galactic-field stars, although their implied injection rate is 1.5 - 2 times less than that of the Galactic tide\footnote{If we do not consider very close encounters (which can occur because the process is stochastic) occurring on very large time scales, probably leading to comet showers (\citealp{Hil81}).} (\citealp{HT86}), have their own important role -- they keep the OC isotropic. The trajectories with ``course inner solar system'' would quickly be depleted if there were no passing stars. Synergy between the Galactic tide and the passing stars ensures almost steady flow of new comets into the inner solar system (\citealp{Ric+08}). Thus all above-mentioned dynamical agents are important in the delivery process.

\subsection{Puzzles}\label{puzzles}
Here we briefly review some of the persistent puzzles that challenge the classical OC theory.

Simulations of OC formation indicate that only 1 - 3 \% of all bodies that are scattered by the giant planets are trapped to the present day outer OC orbits (or $\sim 5 \%$ into the whole cloud; \citealp{Don+04, Kai+11}). This low trapping efficiency leads to some inconsistencies in the standard theory, if we presume that the outer OC is the source of the observed LPCs. Specifically, the primordial protoplanetary disk of planetesimals of the total mass 70 - 300 $M_{\oplus}$ is required to explain the observed LPC flux near Earth. Such a massive disk is at odds with giant planets formation theory, leading to their excessive migration and/or formation of additional giant planets (\citealp{Don+04} and references therein). 

The existence of the mentioned special dynamical pathway described in \citet{KQ09} could serve as a possible solution to this problem, because the trapping efficiency of the inner OC can be an order of magnitude larger than that in the outer OC if the OC formation began in an open cluster (\citealp{KQ08}). In any case, the Sun was more probably born in an embedded cluster (\citealp{LL03}), encased in interstellar gas and dust. The sketched simple solution could be problematic in the presence of a vast amount of gas as in the embedded cluster environment. Aerodynamic gas drag on planetesimals prevents kilometre-sized bodies from entering the cloud and, in the most extreme case, this first stage of the solar system evolution does not make any contribution to the cloud (\citealp{Bra+07}).

Another outstanding puzzle concerns the observed population ratio between the OC and the scattered disk\footnote{It is believed that the scattered disk is the source region of the Jupiter-family comets (\citealp{DL97}).} (SD). Observations suggest that this ratio lies between 100 and 1000 but simulations that produce these two reservoirs simultaneously, yield the value of the order of 10 \citep{DL97, Lev+08, KQ09}. The populations are inferred from the observed fluxes of new LPCs and Jupiter-family comets (JFCs), which are brighter than some reference total magnitude. However, the population ratio  estimated in the simulations of the OC and SD formation refers to objects larger than a given size. Accounting for the fact that ``an LPC is smaller than a JFC with the same total absolute magnitude'', \citet{BM13} arrive at the discrepancy of a factor of ``only'' 4.

As early as the first numerical simulations of the OC formation were performed, it was recognized that only bodies with semi-major axes $a$ beyond $\sim2000$ au could have their perihelia torqued out of the planetary zone into the OC (\citealp{Dun+87}). Bodies with smaller $a$ would still have their perihelia settled near planets. The observed orbital distribution of trans-Neptunian objects (TNOs) have largely agreed with this result. In any case, two striking exceptions have been found - the orbit of Sedna (\citealp{Bro+04}) and 2012 VP$_{113}$ (\citealp{TS14}). With perihelia $(q)$ of 76 and 80 au respectively, these objects no longer interact with planets, yet their large semi-major axes of $\sim 500$ and $\sim 250$ respectively, point to strong planetary perturbations in the past. Although their semi-major axes are larger than most TNOs, they are still too small to be significantly perturbed by the current local Galactic tide. Thus, these orbits remain unexplained by any known dynamical process in the solar system (\citealp{ML04}). An interesting solution to this problem was offered by \citet{Kai+11}, namely, radial migration of the Sun (\citealp{SB02}), which has not been accounted properly in any past study. The simulation of \citet{Kai+11} began with the formation of the Galaxy in a large N-body + smooth-particle-hydrodynamics simulation where solar analogues were identified. Then the OC formation around these stars (often substantial radial migrants) were followed under the influence of the four giant planets, the Galaxy and randomly passing stars, leading to the conclusion that Sedna can be a classical OC body. Unfortunately, the enhanced tidal field that is due to the Sun's radial migration (inward with respect to its current position, if we are looking back in time) also enhances erosion of the outer OC, and thus deepens the primordial disk-mass problem \citep{Kai+11}.

\subsection{Basics of MOND}

According to the MOND\footnote{From now on, when we write ``MOND'' we mean the 1983 Milgrom's formulation \citep{Mil83b}, the simple formula in Eq. (\ref{a1}). When we write ``Milgromian dynamics (MD)'' we mean general theory, like in \citet{BM84} on the classical level or in \citet{Bek04} on the Lorentz-covariant level.} algorithm (\citealp{Mil83b}), the true gravitational acceleration in spherically symetric systems has to be calculated as
\begin{eqnarray}\label{a1}
{\bf g}~=~\nu(g^{N}/a_{0})~{\bf g^{N}}~,
\end{eqnarray}
where $a_{0}\approx10^{-10}$ m s$^{-2}\sim c~H_{0}\sim c^{2}~\Lambda^{1/2}$ is the transition acceleration, $c$ is the speed of light, $H_{0}$ is the Hubble constant, $\Lambda$ is the cosmological constant, ${\bf g^{N}}$ is the expected Newtonian acceleration, $|{\bf g^{N}}|\equiv g^{N}$, and $\nu(\beta)$ is an interpolating function that reflects the underlying general theory with properties $\nu(\beta)\rightarrow 1$ for $\beta\gg1$ and $\nu(\beta)\rightarrow \beta^{-1/2}$ for $\beta\ll1$. Eq. (\ref{a1}) implies that
\begin{eqnarray}\label{a10}
\vert{\bf g}\vert\equiv g=(g^{N}a_{0})^{1/2}~~\Leftrightarrow~~g^{N}\ll a_{0}~,
\end{eqnarray}
and thus it yields exactly the well-known scaling relations \citep{McG+00, FJ76, Mil83a}.
The basics of MOND in Eq. (\ref{a1}) can be written equivalently in the form
\begin{eqnarray}\label{a11}
\mu(g/a_{0})~{\bf g}~=~{\bf g^{N}}~,
\end{eqnarray}
where $\mu(\alpha)=1/\nu(\beta)$, $\beta=\alpha~\mu(\alpha)$, satisfies $\mu(\alpha)\rightarrow 1$ for $\alpha\gg1$ and $\mu(\alpha)\rightarrow \alpha$ for $\alpha\ll1$.
Eq. (\ref{a10}), the backbone of MOND/MD, is the equivalent of stating that: (i) equations of motion are invariant under transformation $(t,{\bf r})\rightarrow(\lambda t,\lambda {\bf r})$, $\lambda\in\mathbb{R}$ \citep{Mil09b}, or (ii) the gravitational field is enhanced by anti-screening of ordinary masses in some gravitationally polarizable medium that is characterized by ``gravitational permittivity'' equal to $g/a_{0}$ \citep{BLT08, BLT09, BB14}. Eventually, MOND can be related to quantum-mechanical processes in the vacuum \citep{Mil99}. Another interesting theory taking the best of both worlds of MD and $\Lambda$CDM, is the recent DM superfluid model \citep{BK15b, bk15a}.

As MD has higher predictive power in galaxies than the $\Lambda$CDM model, although its parent (generally-covariant) theory is still missing, and as most of the classical OC lies in the MD acceleration regime, which is modulated by the external field of the Galaxy $\sim 2a_{0}$, it is asking for the motion of the Oort spike comets to be investigated as it is prescribed by MD. 
Science is mainly about formulating and testing hypotheses.
The possible inevitable tension between the theory and observations could be a disproof of some formulations of MD, incorporating Eq. (\ref{a10}).

Maybe application of the non-standard physics does not yield inconsistencies between the OC formation / OC body-injection models, which are calibrated by the observed LPC flux, and those of giant planets formation, which are calibrated by the appearance of the outer planets region.


\section{Milgromian dynamics}\label{MD}

The simple formula of Eq. (\ref{a1}), when considered as modified gravity\footnote{Eqs. (\ref{a1}) and (\ref{a11}) can be equivalently considered as modified inertia and the whole theory can be built around modifying the kinetic part of the classical action \citep{Mil94, Mil11}. We do not consider modified inertia theories in this paper. Note that these are generically non-local theories \citep{Mil94}.}, cannot be regarded as a universal theory that is applicable to any self-gravitating system of interest, e.g. for not obeying conservation laws out of highly symmetric problems \citep{FM12}. In any case, it was recognised, as early on by \citet{BM84} at the classical level and by \citet{Bek04} at the Lorentz-covariant level, that construction of a universal theory, reproducing Eq. (\ref{a1}) in the special case of the static weak field limit and spherical symmetry, is possible.

\subsection{Quasi-linear formulation of MD}\label{sec:QUMOND}
Several Lorentz covariant theories of MD have been devised in recent years (e.g. \citealp{Bek04, San05, Zlo+07, Mil09a}) which reproduce Eq. (\ref{a1}) in the static weak field limit and spherical symmetry, but differing from each other outside of it \citep{ZF10}. At a classical level these theories generally transform to one of the two types of modified Poisson equation \citep{BM84, Mil10}. Both classical theories are derived from action, thus benefiting from the standard conservation laws.
The theory from \cite{Mil10}, dubbed QUMOND for quasi-linear formulation of MD, can be considered as especially attractive for its computational friendliness.

In QUMOND the field equation that determines MD potential, $\Phi$, reads
\begin{eqnarray}\label{QFE}
\nabla\cdot\left(\nabla\Phi\right)=\nabla\cdot\left[\nu\left(\vert\nabla\phi^{N}\vert/a_{0}\right)\nabla\phi^{N}\right],
\end{eqnarray}
where $\phi^{N}$ is the Newtonian potential fulfilling $\nabla\cdot(\nabla\phi^{N})=4\pi G \varrho_{b}$, $\varrho_{b}$ is baryonic mass density. QUMOND comes from modifying only the gravitational part of the classical action hence the equation of motion stays the same
\begin{eqnarray}\label{eqm}
{\bf g}~=~-\nabla\Phi~.
\end{eqnarray}

Let us define the so-called phantom matter density (PMD)
\begin{eqnarray}\label{roph}
\varrho_{ph}~=~\frac{\nabla\cdot[\widetilde{\nu}(|\nabla\phi^{N}|/a_{0})\nabla\phi^{N}]}{4\pi G},
\end{eqnarray}
$\widetilde{\nu}(\beta)\equiv\nu(\beta)-1$. Eq. (\ref{roph}) does not represent any real physical quantity, particle, or field. PMD is only a mathematical object that allows us to take advantage of the already mentioned QUMOND formulation of MD and write the equations in our intuitive Newtonian sense with ``dark matter''. 
With aid of Eq. (\ref{roph}), the MD potential $\Phi$ can be written as a sum 
\begin{eqnarray}\label{MDphisum}
\Phi~=~\phi^{N}+\phi_{ph}~,
\end{eqnarray}
where the phantom potential $\phi_{ph}$ fulfils normal Poisson equation
\begin{eqnarray}\label{phiph}
\nabla\cdot\left(\nabla\phi_{ph}\right)~=~4\pi G \varrho_{ph}~.
\end{eqnarray}
Once the Newtonian potential is specified, PMD can be found and hence the motion in MD can be traced.

The widely used family of $\widetilde{\nu}(\beta)$ functions, corresponding to the special behaviour of $\nu(\beta)$ in Eq. (\ref{a1}), is
\begin{eqnarray}\label{A3}
\widetilde{\nu}_{n}(\beta)~=~\left[\frac{1+\left(1+4\beta^{-n}\right)^{1/2}}{2}\right]^{1/n}-1~,
\end{eqnarray}
see, e.g. \citet{FM12}.
It is well known that the simple $n=1$ function \citep{FB05} reproduces the rotation curves of the most spiral galaxies well, e.g. \citet{Gen+11}.
However, this function is because of its rather gradual transition to the Newtonian regime excluded by solar system tests, e.g. \citet{SJ06}, \citet{BN11}. 
It is possible to construct an interpolating function with more rapid transition to the Newtonian regime (less impact on the solar system) and, at the same time, very similar to the simple interpolating function on the galactic scales where accelerations are $\sim a_{0}$ (see Fig. 19 in \citealp{FM12}). An example of this is \cite{McG08}:
\begin{eqnarray}\label{expinterpol}
\widetilde{\nu}(\beta)~=~\left(1-e^{-\beta^{1/2}}\right)^{-1}-1~.
\end{eqnarray}
Unless stated otherwise, we use this function throughout the paper, together with the standard value $a_{0}=1.2\times10^{-10}$ m s$^{-2}$= 3700 km$^{2}$ s$^{-2}$ kpc$^{-1}$ \citep{Beg+91, Gen+11, FM12}. 

MD greatly reduces the missing mass in galaxy clusters but leaves consistent mass discrepancy of a factor of about 2 (e.g \citealp{San03}, see also \citealp{FM12}). This fact is frequently used as a reason to completely refute any consideration of MD\footnote{The short argumentation of MD sceptics often goes as ``the Bullet cluster''. In MD theories the mass discrepancies are uniquely predicted by the distribution of baryons but do not need to follow the distribution of baryons exactly.}. There is a suggestion to avoid the remaining discrepancies with a variation of $a_{0}$, and that $a_{0}$ is larger in clusters than it is in galaxies (e.g., \citealp{ZF12, Kho15}). We do not develop this idea in this paper.
In MD, the remaining missing mass does not need to be non-baryonic. Instructed by the history and motivated by the missing baryons problem\footnote{$\sim 30\%$ of the baryons predicted by the big bang nucleosynthesis were not yet detected. Only a fraction of these hidden baryons would be necessary to account for the mass discrepancy in galaxy clusters in MD \citep{FM12}.} it is completely possible that we still do not know the whole baryonic budget of galaxy clusters. The recent discovery of more than a thousand ultra-diffuse galaxy-like objects in the Coma cluster \citep{Kod+15} further promotes this suggestion \citep{Mil15}.

\subsection{Solving for the Milgromian potential of the Galaxy on a grid}\label{PMD}
One can convert known baryonic matter distribution to QUMOND potential and hence the real acceleration. But in general this has to be done numerically. 
According to the scheme sketched in Eqs. (\ref{eqm}) - (\ref{phiph}) first we have to know the Newtonian potential $\phi^{N}({\bf r})$, thus we have to solve the Poisson equation $\Delta\phi^{N}({\bf r})=4\pi G \varrho_{b}({\bf r})$, where the baryonic mass density $\varrho_{b}({\bf r})$ is specified by the adopted model of the Galaxy, see Sect. \ref{Galaxymodel}. For this purpose, we employ a fast Poisson solver on a cartesian grid with the boundary condition that corresponds to a point mass, $\phi^{N}(r)=-GM_{b}/r$, on the last grid point, where $r$ is the centre of mass distance of the baryonic mass density grid and $M_{b}$ is the total baryonic mass.

For a given Newtonian potential $\phi^{N}$ discretised on a cartesian grid $(x,~y,~z)$ of step $h$, the discretised version of Eq. (\ref{roph}) is given on a grid point $(i,~j,~k)$ by:
\begin{eqnarray}\label{A4}
\varrho_{ph}^{i,j,k}~=~\frac{1}{4\pi G h^{2}}&\Big{[}&
   \left(\phi^{N}_{~i+1,j,k}-\phi^{N}_{~i,j,k}\right)\widetilde{\nu}_{B_{x}} \nonumber \\
&-&\left(\phi^{N}_{~i,j,k}-\phi^{N}_{~i-1,j,k}\right)\widetilde{\nu}_{A_{x}}\nonumber \\
&+&\left(\phi^{N}_{~i,j+1,k}-\phi^{N}_{~i,j,k}\right)\widetilde{\nu}_{B_{y}}\nonumber \\
&-&\left(\phi^{N}_{~i,j,k}-\phi^{N}_{~i,j-1,k}\right)\widetilde{\nu}_{A_{y}}\nonumber \\
&+&\left(\phi^{N}_{~i,j,k+1}-\phi^{N}_{~i,j,k}\right)\widetilde{\nu}_{B_{z}}\nonumber \\
&-&\left(\phi^{N}_{~i,j,k}-\phi^{N}_{~i,j,k-1}\right)\widetilde{\nu}_{A_{z}}~\Big{]}~,
\end{eqnarray}
where $\widetilde{\nu}$ function is evaluated in a particular midpoint, e.g. $\widetilde{\nu}_{B_{x}}$ is evaluated in $(i+1/2,~j,~k)$, $\widetilde{\nu}_{A_{y}}$ in $(i,~j-1/2,~k)$, and so on, half a cell from $(i,j,k)$ in each of the three orthogonal directions, see, e.g. \citet{FM12,Lug+13,Lug+14,Lug+15} for illustration. The gradient of $\phi^{N}$ in $\widetilde{\nu}_{B_{x}}(|\nabla\phi^{N}|/a_{0})$ is approximated by $\nabla\phi^{N}=(4\phi^{N}_{~i+1,j,k}-4\phi^{N}_{~i,j,k}~,~\phi^{N}_{~i+1,j+1,k}-\phi^{N}_{~i+1,j-1,k}+\phi^{N}_{~i,j+1,k}-
\phi^{N}_{~i,j-1,k}~,~\phi^{N}_{~i,j,k+1}-\phi^{N}_{~i,j,k-1}
+\phi^{N}_{~i+1,j,k+1}-\phi^{N}_{~i+1,j,k-1})/(4h)$, and so forth.

Finally, knowing the PMD we can solve for the effective Milgromian potential $\Phi({\bf r})$ in $\Delta\Phi({\bf r})=4\pi G [\varrho_{b}({\bf r})+\varrho_{ph}({\bf r})]$ on the same grid. As the boundary condition
\begin{eqnarray}\label{A5}
\Phi(r)=(G M_{b}a_{0})^{1/2}\ln(r)~,
\end{eqnarray}
where $r$ is the centre of mass distance of the ``mass density'' grid and $M_{b}$ is the total baryonic mass, is assumed on the last grid point, in accordance with Eq. (\ref{a1}). In the whole procedure of obtaining $\Phi$, we assume that the Galaxy is isolated from external gravitational fields\footnote{To avoid confusion, we treat the Galaxy as being isolated but we consider the solar system as being embedded in the field of the Galaxy.}, see Sect. \ref{Sec:EFE} for a discussion on EFE. This is a good approximation until the internal gravity becomes comparable with the external field generated by the large scale structure, which is of the order of $a_{0}/100$ \citep{Fam+07}. At the position of the Sun the internal gravity is $\sim a_{0}$.

\subsection{External field effect}\label{Sec:EFE}
A special feature of MD as modified gravity is that its formulation breaks the strong equivalence principle \citep{Mil86b}.

If we have a system $s$ that rests in the gravitational field of a larger system $S$. Say that $S$ generates gravitational acceleration ${\bf g_{e}}=-\nabla\Phi_{e}$ within $s$. 
We assume that the gravitational field that is acting on a body within $s$, ${\bf g}=-\nabla\Phi$, can be separated into internal ${\bf g_{i}}=-\nabla\Phi_{i}$ ($\vert{\bf g_{i}}\vert\equiv g_{i}$) and external ${\bf g_{e}}=-\nabla\Phi_{e}$ ($\vert{\bf g_{e}}\vert\equiv g_{e}$) part. We can then substitute $\nabla\phi^{N}=\nabla\phi^{N}_{~i}+\nabla\phi^{N}_{~e}=-{\bf g^{N}_{i}}-{\bf g^{N}_{e}}$ into Eq. (\ref{QFE}), where ${\bf g^{N}_{i}}$ ($\vert{\bf g^{N}_{i}}\vert\equiv g^{N}_{i}$) and ${\bf g^{N}_{e}}$ ($\vert{\bf g^{N}_{e}}\vert\equiv g^{N}_{e}$) are internal and external Newtonian gravitational accelerations. After removing divergences, dropping the curl-field and considering only directions in the plane perpendicular to the external field this gives \citep{Ang+14}
\begin{eqnarray}\label{simplealgebra}
{\bf g_{i}}=\nu\left(\frac{\sqrt{\left(g^{N}_{i}\right)^{2}+\left(g^{N}_{e}\right)^{2}}}{a_{0}}\right){\bf g^{N}_{i}}~,
\end{eqnarray}
where we have further assumed ${\bf g_{e}}=\nu(g^{N}_{e}/a_{0}){\bf g^{N}_{e}}$.
The internal gravity in $s$ depends not only on internal gravitational sources (in our case - the Sun) but also on the strength of the external field at the position of $s$ (in our case - the local strength of the Galactic gravitational field), even when the external field is considered as being constant within $s$.

This effect should not be confused with tidal forces that arise from the non-uniformity of the external gravitational field across the system $s$. A person in the (arbitrarily small) falling elevator in $s$ can find out about the existence and properties of the external gravitational field through its influence on the internal dynamics. Say $g^{N}_{e}$ is constant, if $g^{N}_{i} < a_{0} \ll g^{N}_{e}$ in Eq. (\ref{simplealgebra}) the system $s$ behaves purely as Newton said, with no sign of the modified dynamics as $\nu(g^{N}/a_{0})$ tends to 1 then, similarly as in the case $g^{N}_{i} \gg a_{0}$. The opposite deep-MD regime applies when $g^{N}_{e} < g^{N}_{i} \ll a_{0}$. The standard MD effects are observed only when both internal and external gravity are sufficiently small $(\lesssim a_{0})$ and, moreover, the external field does not dominate over the internal one.
Eventually, if the hierarchy goes as $g^{N}_{i} < g^{N}_{e} \sim a_{0}$, the dynamics is Newtonian 
with rescaled gravitational constant $G/\mu(g_{e}/a_{0})=\nu(g^{N}_{e}/a_{0})G$, where $G$ is the Newtonian gravitational constant. Moreover, the dynamics is anisotropic with dilatation along the direction of the external field\footnote{This is not seen in approximative Eq. (\ref{simplealgebra}), but see Sect. \ref{MilgromianOC} where a more rigorous approach is applied and anisotropic dynamics emerge.}.

The external field of the Galaxy, ${\bf g_{e}}$, thus has to be considered carefully beyond its tidal effects when modelling MOC. We use the constant value $g_{e}= V^{2}_{0}/R_{0}=240^{2}$ km$^{2}$ s$^{-2}$/(8.3 kpc)$~\doteq1.87~a_{0}$, where $V_{0}$ is the circular speed of the Sun at $R_{0}$, and $R_{0}$ is the distance between the Sun and the Galactic center (GC), throughout the paper. Compare the values of $V_{0}$ and $R_{0}$ with for example those given by \citet{Sch12}. We take the Newtonian value $g^{N}_{e}$ as a solution of
\begin{eqnarray}\label{getogen}
g_{e}=\nu(g^{N}_{e}/a_{0})g^{N}_{e}~.
\end{eqnarray}
Eq. (\ref{getogen}) is known to be a good approximation at the position of the Sun \citep{BM95} (the Galaxy can be well modelled as being made up of bulge plus exponential disks).

We note that the Galactic tide is modelled as a separate effect, see Sect. \ref{tide} for details.

\begin{figure}
\begin{center}
\resizebox{0.85\hsize}{!}{\includegraphics{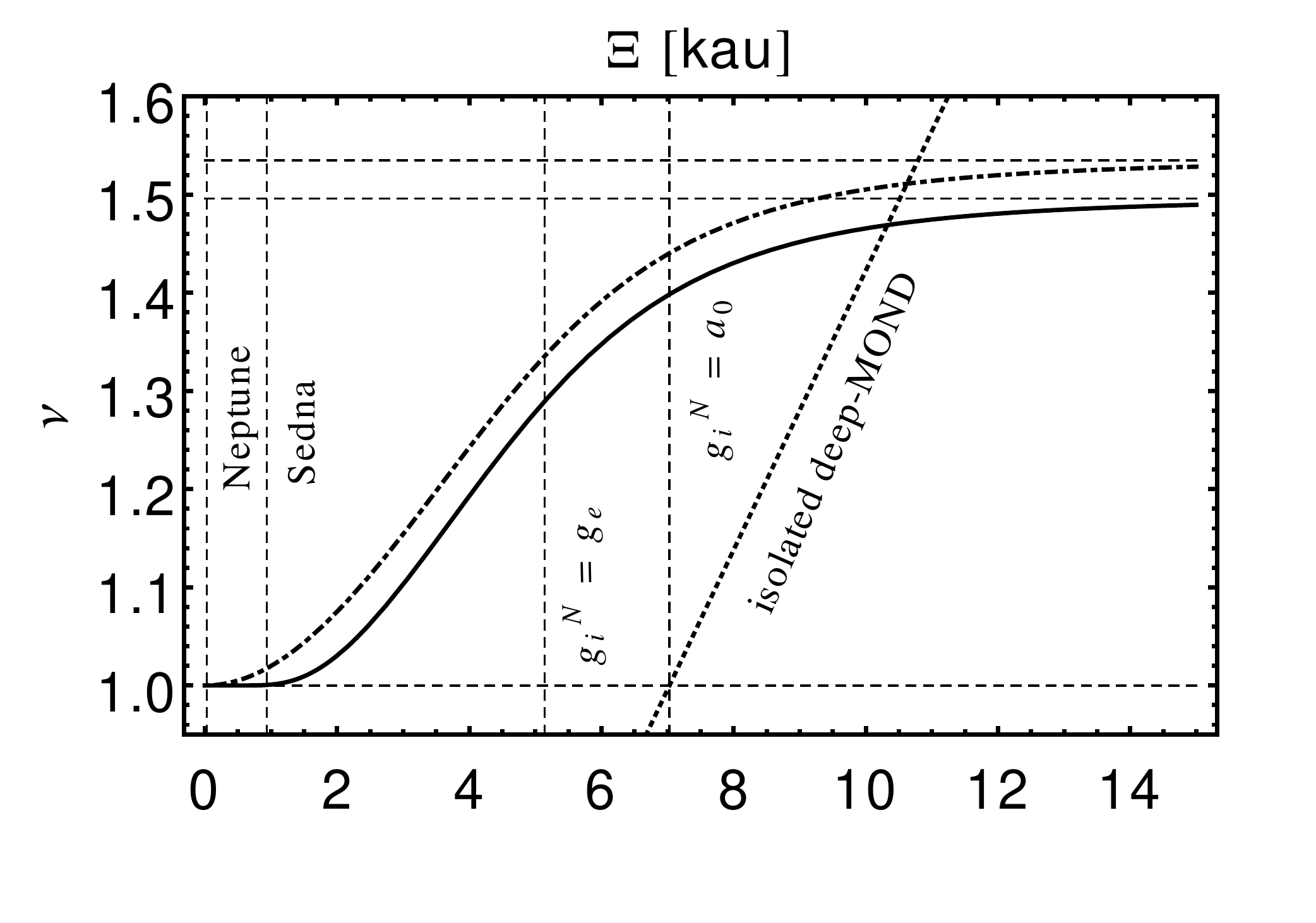}}
\caption{Interpolating functions $\nu(\beta)=[1+(1+4\beta^{-1})^{1/2}]/2$ (dot-dashed line) and $\nu(\beta)=[1-\exp(-\beta^{1/2})]^{-1}$ (solid line) as functions of heliocentric distance $\Xi$.                                                                    $\beta\equiv g_{N}/a_{0}$ is approximated with $[(g^{N}_{e})^{2} + (GM_{\odot}/\Xi^{2})^{2}]^{1/2}/a_{0}$, i.e. vectors of external and internal Newtonian gravity are assumed to be perpendicular to each other for simplicity. The two topmost horizontal dashed lines are the values $\nu$-functions asymptote to under the condition $\Xi\rightarrow\infty$ (then $g_{N}\rightarrow g^{N}_{e}$), the downmost $\nu=1$ marks the Newtonian limit. Vertical dashed lines from left to right indicate the aphelia of Neptune and Sedna and the distances where $GM_{\odot}/\Xi^{2}=g_{e}=1.9 a_{0}$ and $GM_{\odot}/\Xi^{2}=a_{0}$. The dotted line is $\beta^{-1/2}= [(GM_{\odot}/\Xi^{2})/a_{0}]^{-1/2}$, the deep-MOND limit of $\nu(\beta)$, in the case of no external field.}
\label{img:MONDSS}
\end{center}
\end{figure}

To better visualise the gravity-boosting effect of MD and also the importance of EFE on the solar system scales, we plot $\nu$ interpolating function as a function of heliocentric distance $\Xi$ in Fig. \ref{img:MONDSS}. The simple $\nu(\beta)=[1+(1+4\beta^{-1})^{1/2}]/2$ and the exponential $\nu(\beta)=[1-\exp(-\beta^{1/2})]^{-1}$ interpolating functions are depicted.
$\beta\equiv g_{N}/a_{0}$ is approximated with $[(g^{N}_{e})^{2} + (GM_{\odot}/\Xi^{2})^{2}]^{1/2}/a_{0}$, i.e. vectors of external and internal Newtonian gravitational acceleration are assumed to be perpendicular to each other for simplicity. The characteristic distance scale (MD transition scale) is $\sim\sqrt{GM_{\odot}/a_{0}}\approx 7$ kau. Because of the action of EFE, $\nu(\beta)$ does not diverge with $\Xi\rightarrow\infty$, but asymptotes to the constant value $\nu(g^{N}_{e}/a_{0})$.

EFE is important, even in the high-acceleration regime, where the gravity-boosting effect of MD is very weak. It has been shown that, at $\Xi\ll \sqrt{GM_{\odot}/a_{0}}$, which is well fulfilled in the planetary region, EFE manifests primarily through an anomalous quadrupolar correction to the Newtonian potential, which increases with the heliocentric distance $\Xi$ \citep{Mil09c, BN11}. This dynamical effect is thus analogous to that of a massive body hidden at a large heliocentric distance, lying in the direction to GC, ${\bf g_{e}}/g_{e}$, \citep{Hog+91,Ior10b}. As the external field ${\bf g_{e}}$ rotates with period $\sim210$ Myr, this corresponds to an unfeasible configuration in Newtonian dynamics (too massive body in a very distant circular orbit around the Sun). Hence the effect of MD should be distinguishable from that of the distant planet in simulations that are carried out on large timescales.

\section{Models}\label{models}

In Sect. \ref{Galaxymodel}, the adopted model of the Galactic matter distribution is presented and the appropriate PMD for this model is calculated. The model of the Galaxy is considered solely to estimate the matter density in the solar neighborhood and hence estimate the effect of the Galactic tide, see sections \ref{tide}, \ref{simul} and \ref{XXZ}. In Sect. \ref{sec:simpleOC} the simplified model of the MOC that is embedded in a constant external field is introduced. The majority of the qualitative analysis performed in the paper is carried out assuming this simple model. 

Firstly, we erect a rectangular Galilean coordinate system $O_{\odot}(\xi',\eta',\zeta')$ that is centred on the Sun. At time $t=0$ (present time), the inertial reference frame $O_{\odot}(\xi',\eta',\zeta')$ coincides with the rotating Galactic rectangular coordinate system, i.e. $\xi'$ axis is directed from the Sun to the GC at $t=0$. We also use an inertial frame that is centred on the GC, denoted $O_{GC}(x,y,z)$, with $x-y$ plane being the Galactic plane and $x$ axis directed from the GC to the Sun at $t=0$.

\subsection{The Galaxy}\label{Galaxymodel} 

We adopt the Galaxy mass model of \citet{McG08}, similar to that used in \citet{Lug+14}. \citet{McG08} concluded that MOND prefers short disk scale lengths in the range $2.0<r_{d}<2.5$ kpc. The modelled Galaxy consists of a stellar double-exponential disk with the scale length $R_{d}=2.3$ kpc and the scale height $z_{d}=0.3$ kpc with the disk mass $4.2\times10^{10}~M_{\odot}$. Moreover, it has a thin gas disk of the total mass $1.2\times10^{10}~M_{\odot}$ with the same scale length and half scale height as the stellar one and a bulge modelled as a Plummer's sphere, with the mass $0.7\times10^{10}~M_{\odot}$ and the half-mass radius 1 kpc.

\subsubsection{Phantom matter density}\label{pomodoro}

MD predicts the complex structure of a ``Newtonist's dark halo'' with a pure disk component and rounder component with radius-dependent flattening that becomes spherical at great distances \citep{Mil01}, see also Fig. 5 in \citet{Lug+15}.

\begin{figure}\centering
\resizebox{0.5\hsize}{!}{\includegraphics{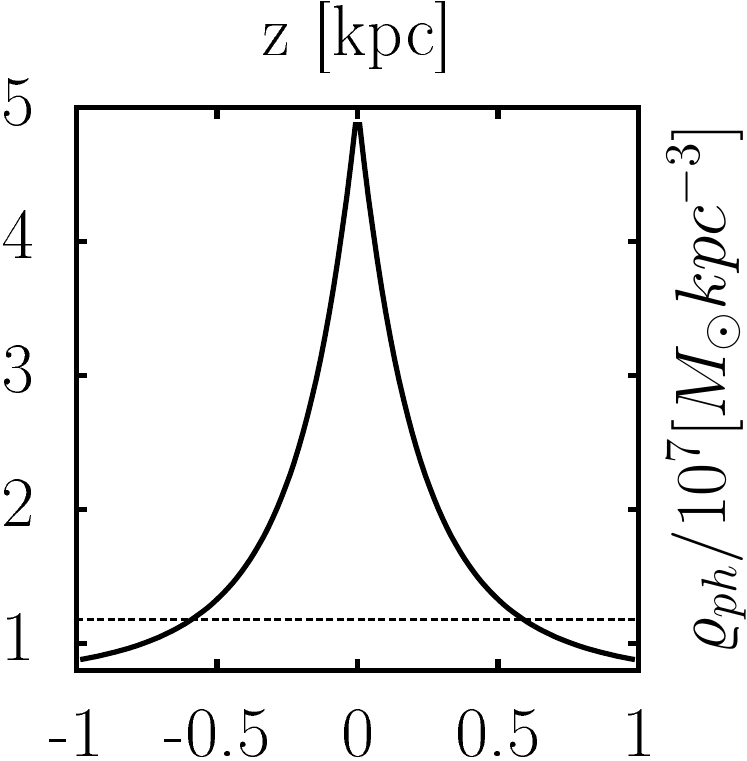}}
\caption{PMD of the Galaxy (solid), modelled as in Sect. \ref{Galaxymodel} at $R=R_{0}=8.3$ kpc within $\vert z \vert<1$ kpc. NFW dark matter density (dashed line) is also depicted.}
\label{img:pmd}
\end{figure}

We calculated the PMD of the Galaxy model according to the numerical scheme of Sect. \ref{PMD}. A cartesian $(x,y,z)$ grid with $512\times512\times256$ cells and resolution of  $0.1\times0.1\times0.02$ kpc was used. This resolution was tested as being sufficiently fine enough so that the calculated PMD changes only negligibly if the resolution is further increased.
Fig. \ref{img:pmd} shows the vertical PMD $\varrho_{ph}(z)$ at $R=R_{0}=8.3$ kpc within $\vert z \vert<1$ kpc.
The $K_{z}$ force perpendicular to the Galactic plane will be obviously enhanced in this case, compared to the Galaxy that resides in a spherical DM halo, as predicted by Milgrom already in his pioneer paper \citep{Mil83b}. 

Owing to small stellar samples (Hipparcos data), one cannot precisely recover the shape of $K_{z}(z)$ or of the dynamical density, only the surface density below some $\vert \overline{z} \vert$, where $\overline{z}$ is the  mean  distance  of  the  samples  from  the  Galactic
plane \citep{Bie+09}. 
We should compare the calculated surface density of the baryonic matter plus the phantom matter with observations. \citet{HF04} find the dynamical surface density $\Sigma_{0}=74\pm 6$ M$_{\odot}$ pc$^{-2}$ within $\vert z \vert<1.1$ kpc.
By fitting the calculated local PMD with a superposition of three exponential disks, we find $\Sigma_{0}=80$ M$_{\odot}$ pc$^{-2}$ within $\vert z \vert<1.1$ kpc, which is consistent with the value of \citet{HF04}. The portion 43 M$_{\odot}$ pc$^{-2}$ resides in the normal matter and 37 M$_{\odot}$ pc$^{-2}$ in the phantom.

\subsubsection{The dark matter halo of Newtonian Galaxy}\label{halo}
The Navarro-Frenk-White (NFW) halo model \citep{Nav+97}
\begin{eqnarray}\label{NFW}
\varrho_{h}~=~\frac{\varrho_{h,0}}{\delta(1+\delta)^{2}}~,
\end{eqnarray}
where $\delta\equiv r/r_{h}$, $r_{h}$ is the scale radius (spherically symmetric halo, $r$ is radial coordinate), $\varrho_{h,0}$ is a constant, represents the culmination of the present day theoretical knowledge in the standard CDM-based cosmology. 

In Sect. \ref{XXZ} we aim to compare the effect of the Galaxy on the MOC and the classical OC. 
We use the NFW model as the model of the Galaxy's dark matter halo in the Newtonian framework in order to find local mass density in the solar neighbourhood and quantify the Galactic tide.

CDM haloes are routinely described in terms of their virial mass, $M_{vir}$, which is the mass that is contained within the virial radius $r_{vir}$, and
the concentration parameter $c = r_{-2}/r_{vir}$, where $r_{-2}$ is the radius at which the logarithmic slope of the density profile $d\log \varrho_{h}/$ $d \log r = -2$ (for the
NFW profile, $r_{-2} = r_{h}$ ). The virial radius $r_{vir}$ is defined as the radius
of a sphere that is centred on the halo centre, which has an average density that is $\Delta$ times the critical density $\varrho_{crit}=3H^{2}_{0}/(8\pi G)$, where $H_{0}$ is the Hubble constant. $\Delta$ varies with redshift, with $\Delta\approx100$ today. For the NFW model
\begin{eqnarray}\label{NFW2}
\frac{\varrho_{h,0}}{\varrho_{crit}}~=~\frac{\Delta}{3}~\frac{c^{3}}{\ln(1+c)-c/(1+c)}
\end{eqnarray}
holds. Thus knowing the concentration parameter $c$ we can find $\varrho_{h,0}$ of Eq. (\ref{NFW}). \citet{Boy+10} examined (NFW) haloes taken from the Millennium-II simulations at redshift zero, in the mass range $10^{11.5}$ $\leq$ $M_{vir} [h^{-1} M_{\odot}]$ $\leq$ $10^{12.5}$, a mass
range that the Galaxy's halo is likely to lie in, and determined that the probability distribution of the concentration parameter was well-fitted by a Gaussian distribution in $\ln c$, with $\langle\ln c\rangle=2.56$ and $\sigma_{\ln c}=0.272$. We adopt $c=\exp(2.56)$ as the concentration parameter of the Galaxy. The remaining degree of freedom in Eq. (\ref{NFW}), represented by the scale radius $r_{h}$, can be eliminated by fitting the circular speed $V_{0}$ at radial distance $R_{0}$: $V^{2}_{0}=V^{2}_{d,s}+V^{2}_{d,g}+V^{2}_{b}+V^{2}_{h}$, where the added squared speeds represent particular Galactic components (stellar disk, gas disk, bulge, dark halo) determined by the particular masses that are enclosed within $R_{0}$. Doing so for $V_{0}=240$ km/s, $R_{0}=8.3$ kpc we find: $\varrho_{h,0}=5.750\times 10^{6}$ M$_{\odot}$ kpc$^{-3}$, $r_{h}=28.4$ kpc. Surface density of the NFW halo within $\vert z \vert<1.1$ kpc is 26 M$_{\odot}$ pc$^{-2}$, consistent with the lower bound on $\Sigma_{0}$ \citep{HF04}.

\subsubsection{Galactic tide}\label{tide}
\begin{figure*}\centering
\includegraphics[width=0.7\linewidth]{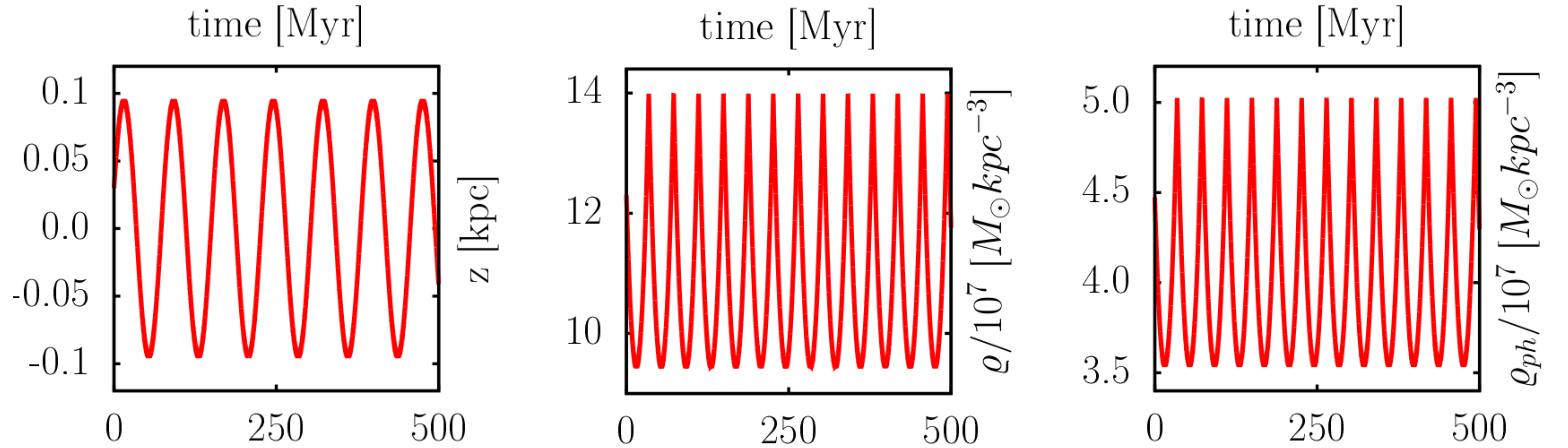}
\caption{\textbf{Left}: Oscilation of the Sun governed by Eq. (\ref{osc}) in MD. We used $z_{\odot}(0)=30$ pc and $v_{z_{\odot}}(0)=7.25$ km s$^{-1}$ as the initial conditions of the Sun's motion. \textbf{Middle}: Local ``total matter density'' $\varrho=\varrho_{b}+\varrho_{ph}$, as experienced by the oscillating Sun. \textbf{Right}: Local PMD, as experienced by the oscillating Sun.}\label{img:sz}
\end{figure*}

We use a 1D model of the Sun's motion through the Galaxy with the Sun moving in a circular orbit upon which are superimposed small vertical oscillations. For the vertical (perpendicular to the Galactic midplane) acceleration of the Sun at $z=z_{\odot}$ we assume
\begin{eqnarray}\label{osc}
\ddot{z}(z_{\odot})~=~-\frac{\partial\Phi}{\partial z}(z_{\odot})~=~-4\pi G \int^{z_{\odot}}_{0}\varrho(z)dz~,
\end{eqnarray}
where in MD, $\varrho(z)=\varrho_{b}(z)+\varrho_{ph}(z)$, is the local vertical ``matter density'' which is sum of the baryonic and the phantom density at $R=R_{0}$ and $\Phi$ is the QUMOND potential of the Galaxy, see sections \ref{PMD} and \ref{Galaxymodel}. In Newtonian dynamics, $\varrho(z)=\varrho_{b}(z)+\varrho_{h}(z)$, where $\varrho_{h}(z)$ is the vertical density of the DM halo at $R=R_{0}$. Eq (\ref{osc}) hangs on the fact that the rotation curve of the Galaxy is approximately flat at the position of the Sun - for an axisimmetric model of the Galaxy: $(1/R)\partial(R\partial\Phi/\partial R)/\partial R+\partial^{2}\Phi/\partial z^{2}=4\pi G \varrho(R,z)$ with $\partial(R\partial\Phi/\partial R)/\partial R\approx0$ holds. Fig. \ref{img:sz} shows the oscillations of the Sun through the Galactic disk governed by Eq. (\ref{osc}). The oscillations have a period of 76.7 Myr. The model of the Galaxy of Sect. \ref{Galaxymodel} is employed.

We approximate the tidal acceleration of a comet\footnote{MD is non-linear. One cannot a priori sum up partial accelerations to get a net acceleration vector. The usage of Eq. (\ref{tidesimple}) in MD is further discussed and justified in Sect. \ref{simul}.} in the inertial frame of reference $O_{\odot}(\xi',\eta',\zeta')$ centred on the Sun as $(0,~0,~\ddot{\zeta'}_{tide}\equiv\ddot{z}_{c}-\ddot{z}_{\odot})$ with
\begin{eqnarray}\label{tidesimple}
\ddot{\zeta'}_{tide}~=~-4\pi G \varrho(z_{\odot})\zeta' + \mathcal{O}(\zeta'^{2})~,
\end{eqnarray}
where $z_{c}$ and $z_{\odot}$ are vertical components (perpendicular to the Galactic midplane) of the position vector of the comet and the Sun with respect to the GC and $z_{c}=z_{\odot}+\zeta'$ holds. 
We omit the $\xi'$ and $\eta'$ components of the tide since these are approximately an order of magnitude smaller than the $\zeta'$ component \citep{HT86}. We note that this is true not only in Newtonian dynamics, but also in MD as the distribution of the phantom matter resembles that of a disk close to the galactic midplane.

\subsection{Simple model of the Milgromian Oort cloud}\label{sec:simpleOC}
Here we introduce a simple model of the MOC embedded in an external field of constant magnitude (no tides). Accounting for the external field is a necessary step as in MD the external field does not decouple from the internal dynamics.

We assume that the Sun travels with angular frequency $\omega_{0}=V_{0}/R_{0}$ in a circular orbit of radius $R_{0}$ which lies in the Galactic midplane $(z=0)$.

Let the Newtonian external field of the Galaxy at the position of the Sun be approximated by the time-dependent vector:
\begin{eqnarray}\label{gesimple}
{\bf g^{N}_{e}}=[g^{N}_{e}~\cos(\omega_{0}t),~g^{N}_{e}~\sin(\omega_{0}t),~0]
\end{eqnarray}
in $O_{\odot}(\xi',\eta',\zeta')$. 
So that at $t=0$: ${\bf g^{N}_{e}}=g^{N}_{e}~{\bf \hat{\xi'}}$, where ${\bf \hat{\xi'}}$ is the unit vector.
In Eq. (\ref{gesimple}), we assume that the Sun orbits counterclockwise in the plane $\xi'-\eta'$ of $O_{\odot}(\xi',\eta',\zeta')$. The sense of rotation of the Sun does not play a role in the analysis.

In Eq. (\ref{roph}) we now have $\nabla\phi^{N}_{S.S.}=GM_{\odot}{\bf \Xi'}/\Xi'^{3}-
{\bf g^{N}_{e}}$, where ${\bf \Xi'}=[\xi',\eta',\zeta']$,
$\Xi'\equiv(\xi'^{2}+\eta'^{2}+\zeta'^{2})^{1/2}$ and the lower index ``S.S.'' stresses that we are dealing with the solar system embedded in the external field of the Galaxy. For the PMD we thus obtain
\begin{eqnarray}\label{rophsimple}
\varrho_{ph,S.S.}~=~\frac{\nabla\widetilde{\nu}\cdot(GM_{\odot}{\bf \Xi'}/\Xi'^{3}-
{\bf g^{N}_{e}})}{4\pi G}~,
\end{eqnarray}
where $\widetilde{\nu}\equiv\widetilde{\nu}(|GM_{\odot}{\bf \Xi'}/\Xi'^{3}-
{\bf g^{N}_{e}}|/a_{0})$. 
The phantom potential $\phi_{ph,S.S.}$ can be found by solving the ordinary Poisson equation
\begin{eqnarray}\label{phpot}
\Delta\phi_{ph,S.S.}=4\pi G \varrho_{ph,S.S.}~,
\end{eqnarray}
with the boundary condition: $\phi_{ph,S.S.}=-{\bf g_{e}}\cdot{\bf \Xi'}$.
The equation of motion in $O_{\odot}(\xi',\eta',\zeta')$ then reads
\begin{eqnarray}\label{eqmo}
{\bf \ddot{\Xi'}}=-\nabla\Phi_{S.S.}-{\bf g_{e}}~,
\end{eqnarray}
where $\Phi_{S.S.}=-GM_{\odot}/\Xi'+\phi_{ph,S.S.}$.

As QUMOND equations are linear when formulated with the aid of phantom matter, we can also look for a solution of Eq. (\ref{phpot}) with the vacuum boundary condition ($\phi_{ph,S.S.}=0$ at the boundary) and then evolve a body with
\begin{eqnarray}\label{eqmo2}
{\bf \ddot{\Xi'}}=-\nabla\Phi_{S.S.}~.
\end{eqnarray}

\subsubsection{Simple model of the Oort cloud - numerical solution at t=0}
For integration of cometary orbits throughout the paper we employ the well-tested RA15 routine \citep{Eve85} as part of the {\small MERCURY 6} gravitational dynamics software package \citep{Cha99}, which we have modified appropriately to be compatible with the MD framework. Eq. (\ref{gesimple}) has to be transformed from $O_{\odot}(\xi',\eta',\zeta')$ to a coordinate system used by {\small MERCURY 6}. This transformation and subsequent modification of Eqs. (\ref{rophsimple}) and (\ref{eqmo}) are straightforward. $O_{\odot}(\xi,\eta,\zeta)$ denotes from now on the rectangular coordinate system we use in {\small MERCURY 6}, i.e. Galilean coordinates coinciding at $t=0$ with the heliocentric ecliptical coordinate system\footnote{$O_{\odot}(\xi',\eta',\zeta')$ vs. $O_{\odot}(\xi,\eta,\zeta)$, primed are Galactic and non-primed are ecliptic coordinates at $t=0$.}.

During short time periods, compared to the period of the Sun's revolution around the GC, $\sim210$ Myr, one can approximate Eq. (\ref{gesimple}) with the constant vector ${\bf g^{N}_{e}}=[g^{N}_{e},~0,~0]$~, $g^{N}_{e}\approx 1.22~a_{0}$, in $O_{\odot}(\xi',\eta',\zeta')$.
We used this approximation to find the phantom potential $\phi_{ph,S.S.}$ experienced by a body in the MOC model that is represented by Eqs. (\ref{gesimple}) - (\ref{eqmo}). The numerical procedure is analogous to the one described in Sect. \ref{PMD}. The boundary conditions are described under Eq. (\ref{phpot}).
We employed a regular cartesian grid with $512^{3}$ cells and resolution of 390 au that is centred on the Sun. This resolution was tested to be sufficiently fine enough so that the trajectories of comets do not change significantly if the resolution is further increased. In the case of inner OC orbits in sections \ref{Sedna} and \ref{Sedna2} we used a resolution of 78 au with the same result. The calculated phantom acceleration, $-\nabla\phi_{ph,S.S.}$, is linearly interpolated to an instaneous position of the body within each integration cycle.
We refer to this simplified dynamical model of the MOC as ``simple model of the MOC''.

\subsection{Escape speed}
An isolated point mass $M$ at distance $r\gg(GM/a_{0})^{1/2}$ is in MD source of the potential of the form
\begin{eqnarray}\label{A55}
\Phi(r)\sim (GMa_{0})^{1/2}\ln(r)~.
\end{eqnarray}
Eq. (\ref{A55}) yields asymptotically flat rotation curves but also means that there is no escape from the central field produced by the isolated point mass in MD, since $V^{2}_{esc}(r)\sim\Phi(\infty)-\Phi(r)$. But, an external field (which is always intrinsically present) actually regularizes the former divergent potential, so that it is possible to escape from non-isolated point masses in MD \citep{Fam+07}, as we have already seen in Sect. \ref{Sec:EFE}. 

The escape speed of a comet can be well defined as \citep{Wu+07, Wu+08}
\begin{eqnarray}\label{vesc1}
V_{esc}(\xi,\eta,\zeta)=\sqrt{-2\Phi_{i}(\xi,\eta,\zeta)}~,
\end{eqnarray}
with $-\nabla\Phi_{i}={\bf \ddot{\Xi}}$.
The estimate of the escape in the direction perpendicular to the external field can be found
by approximating the Galactic EFE that is acting on the OC with the simple curl-free formula of Eq. (\ref{simplealgebra}), where now ${\bf g_{i}^{N}}=-GM_{\odot}{\bf \Xi}/\Xi^{3}$.
For the escape speed at $\Xi=r_{C}$, we then have
\begin{eqnarray}\label{vesc}
V_{esc}(r_{c})=\left[2\int^{\infty}_{r_{c}}g_{i}(\Xi)d\Xi\right]^{1/2}~, 
\end{eqnarray}
where $g_{i}(\Xi)\equiv \vert{\bf \ddot{\Xi}}\vert$. We use Eq. (\ref{vesc}) in sections \ref{escapingcomets} and \ref{JSBinMOND} to estimate binding energy of a comet.

\begin{figure*}
\begin{center}
\resizebox{0.65\hsize}{!}{\includegraphics{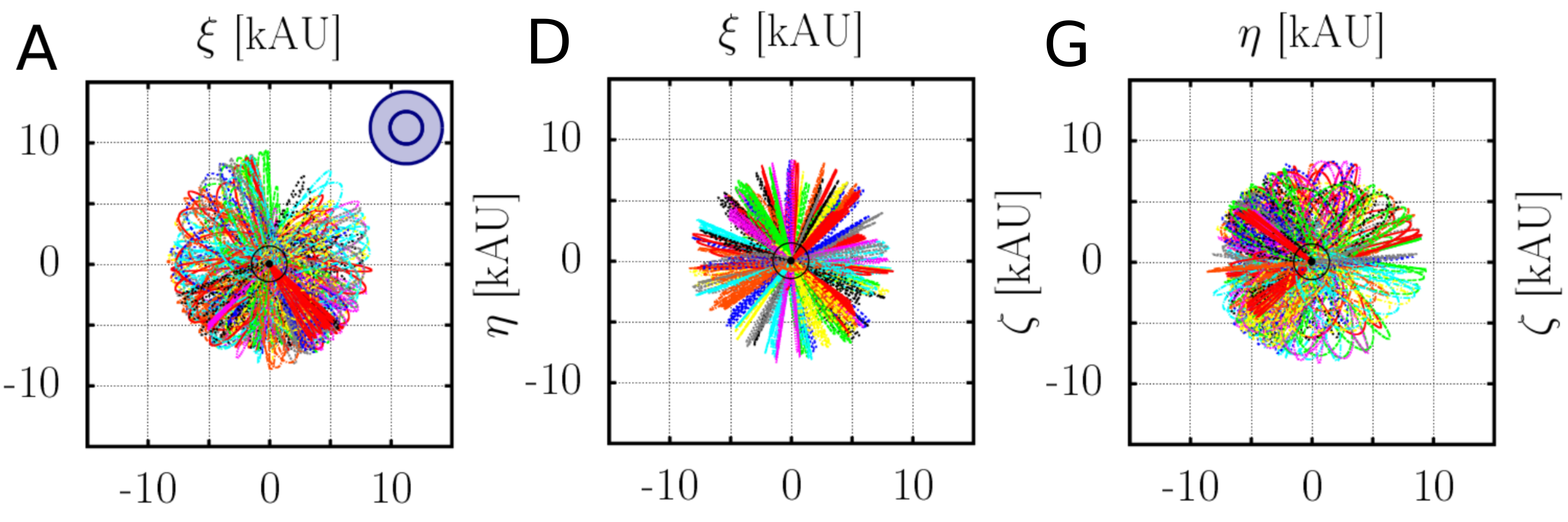}}
\resizebox{0.65\hsize}{!}{\includegraphics{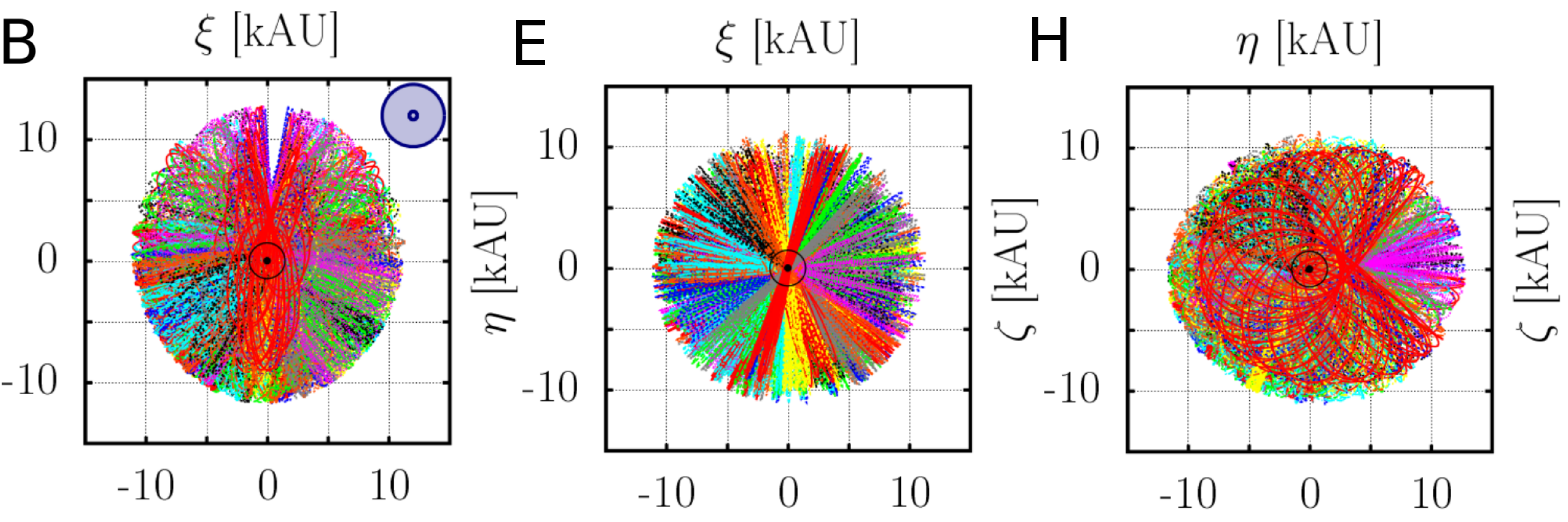}}
\resizebox{0.65\hsize}{!}{\includegraphics{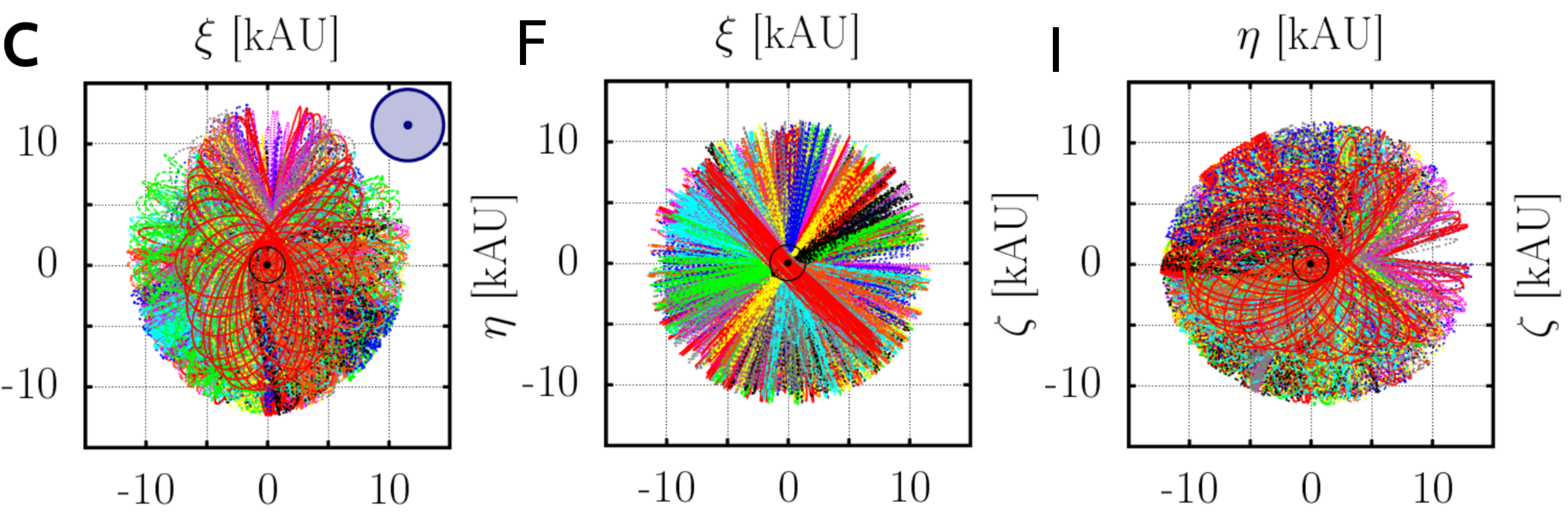}}
\caption{Past Milgromian trajectories of $3\times 100$ Monte Carlo particles projected to 3 mutually orthogonal planes of $O_{\odot}(\xi,\eta,\zeta)$. The particles were initialised with original Newtonian orbital elements: $a=10$ (\textbf{top row}), 50 (\textbf{middle row}), 100 (\textbf{bottom row}) kau, $q$ distributed uniformly on the interval $(0,8)$ au, $\cos(i)$ distributed uniformly on the interval $(-1,1)$, $\omega$ and $\Omega$ distributed uniformly on the interval $(0,2\pi)$, among the particles, and mean anomaly $M=0$. Then the particles were evolved back in time in the simple model of the MOC, Sect. \ref{sec:simpleOC}, for one Keplerian period ($\approx$ 1 Myr) in the case of $a=10$ kau, and for 10 Myr in the case of $a=50$ and 100 kau. The concentric circles at the top right corner of figures A,B, and C represent relative radii of the Milgromian (always the smaller circle) and Newtonian OC (radius $=2a$; always the larger circle) as determined by the simulation and assuming that the cloud is the smallest sphere encompassing all orbits of given initial $a$. 
The Sun resides at [0,0], as indicated by the symbol.}
\label{img:oo}
\end{center}
\end{figure*}

\section{The Oort cloud as seen by a Milgromian astronomer}\label{MilgromianOC}
Do the observations lead us to hypothesize the existence of a vast cloud of bodies as a reservoir of new comets also if we interpret the data with the laws of MD? If so, how vast and shaped, in a rough sense, would be the cloud, compared to the classical one?

\citetalias{DK11} studied the dynamical evolution of 64 Oort spike comets with orbits what were determined with the highest precision, discovered after 1970, with their original semi-major axes larger than 10 kau and osculating perihelion distances $q>3$ au (to minimise non-gravitational effects). They identified 31 comets as dynamically new (having their first approach to the zone of significant planetary perturbations; for the detailed definition see the paper), and one of these comets as possibly hyperbolic\footnote{\citetalias{DK11} found the original reciprocal semi-major axis of the comet C/1978 G2 to be $-22.4\pm 37.8~\times 10^{-6}$ au$^{-1}$.}. Median value of the original reciprocal semi-major axis for the 30 comets on the certainly bound orbits is 22.385 $\times$ 10$^{-6}$ au$^{-1}$, which corresponds to 44.7 kau, maximum and minimum values in the sample read 250.6 and 21.9 kau, respectively. All the orbits have osculating $q<9$ au. The orbits of dynamically new comets are free from planetary perturbations and can be used to study the source region of these comets. We emphasize that for a comet being dynamically new under Newtonian dynamics does not necessary mean to be dynamically new under MD. A reconsideration of the dynamical status in MD would require a similar approach as in \citetalias{DK11} with an extensive use of orbital clones to cover the large errors that are in the original orbital energy. 

To acquire vital motivation we have used a more straightforward approach as a first step. Employing the aforementioned simple model of the MOC, we traced the past trajectories of 300 Monte Carlo test particles that represent a sample of Oort spike comets. We consider this a fairly small sample since, in reality, observed samples are of similar or even smaller numbers. We considered three values of the particle's initial semi-major axis $a=$ 10, 50 and 100 kau. For each of the three values of $a$ we initialised $100$ test particles at their perihelia - all the perihelia lie in the deep Newtonian regime - with the following randomly generated original Newtonian orbital elements: $q$ distributed uniformly on the interval $(0,8)$ au, $\cos(i)$ distributed uniformly on the interval $(-1,1)$, $\omega$ and $\Omega$ distributed uniformly on the interval $(0,2\pi)$, among the test particles, here $q$ is perihelion distance, $i$ is inclination with respect to the ecliptics, $\omega$ is argument of periapsis, and $\Omega$ is the longitude of the ascending node.
The initial Newtonian orbital elements are immediately transformed into the initial cartesian positions and velocities, the notions being independent of the dynamical framework; also these are the observables on the basis of which the orbital elements are calculated\footnote{Published catalogues and papers usually offer only the Newtonian orbital elements, not the observables.}.
We followed the particles with $a=10$ kau back in time for one Keplerian period (which is by no means the real period assuming MD), $2\pi$ ($a$[au])$^{3/2}$/$k$ days, where $k$ is Gaussian gravitational constant, and the particles with $a=50$ and 100 kau for 10 Myr. We do not use the integration time of one Keplerian period in the latter case because, during this time, the change in the external field direction cannot be neglected ($a=100$ kau orbit has the Keplerian period $T_{Kep}\approx 32$ Myr). In any case, as will be shown, all the particles with initial $a=50$ and 100 kau revolve many times during 10 Myr.

By the term ``original orbit'' we want to emphasise the fact that, in reality, the outer planets and non-gravitational effects are important dynamical agents, primarily changing the value of the semi-major axis. We can imagine the ensemble of the initial orbital elements as the result of backward integration of observed osculating (instantaneous) orbits to the time when the comets/particles enter the planetary zone.

The past QUMOND trajectories of the particles are shown in Fig. \ref{img:oo}. Trajectories can be typically described as ellipses with a quickly precessed line of apsides. Moreover, the external field often changes perihelion distances of the particles rapidly and almost irrespective of their initial semi-major axis. This important fact is discussed in Sect. \ref{XXZ}. In this case, the orbits change their shape dramatically, as was previously illustrated in \citet{Ior10} for the deep-MD orbits only.

A small departure from the isotropy of the cloud can be seen in Fig. \ref{img:oo}. The cloud is prolonged in the direction of the $\eta$ axis. Also an indistinct pac-man shape of $\xi-\eta$ and $\eta-\zeta$ plane cuts emerges. This is because of the external field of the Galaxy, which points in the direction of $-{\bf\hat{x}}$ of $O_{GC}(x,y,z)$ (the direction Sun-GC at $t=0$), which also approximately corresponds to the direction of -\boldmath${\hat{\eta}}$\unboldmath~ of $O_{\odot}(\xi,\eta,\zeta)$. The gravity is stronger at negative $\eta$ than at positive. This can be most easily noticed on the $\nu(\beta)$ dependence on the vector sum in the grossly approximative formula ${\bf g_{i}}=\nu(\vert{\bf g^{N}_{i}} + {\bf g^{N}_{e}}\vert /a_{0}){\bf g^{N}_{i}}$ (note that larger $\beta$ means smaller $\nu(\beta)$). We also note the smaller precession rate of the projected orbits in $\xi-\zeta$ plane. Again, this is because the $\xi$ and $\zeta$ components of the Galactic external field are much smaller than the $\eta$ component. 

In any case, the most important result is that even the orbits with initial $a=100$ kau are confined in a cube of side $\sim 28$ kau. in this case, the Newtonian cube would be of side $\sim 400$ kau. This implies that the OC as revealed by comets with original $0<a<100$ kau and interpreted by MD could be much more compact than the Newtonian one. 

These findings looks problematic for MD at first sight. The classical picture of the Galactic tide, as the most effective comet injector, is that the sufficient decrease in a comet's perihelion distance during one revolution - to be able to penetrate the Jupiter-Saturn barrier - can be made only for comets with $a>20 - 30$ kau (e.g. \citealp{Lev+01, Ric14}), hence the comets with aphelion distances that are larger than 40 - 60 kau, if eccentricity is close to 1. These are much larger heliocentric distances than those of the particles in the MOC simulation. Also, comets of the classical inner OC, which take advantage of the Jupiter-Saturn barrier by inflating their semi-major axes, come through this outer region ($a>20 - 30$ kau; i.e. the comets appear to be from the outer OC) where the final decrease in perihelion distance is effectively made \citep{KQ09}. All these findings are of course Newtonian. The tidal field of the Newtonian Galaxy that is embedded in the DM halo is a little different from the QUMOND one, especially its vertical (perpendicular to the Galactic midplane) part. Moreover, completely beyond the tides, the MD's EFE can have a decisive influence on the dynamics.
We address this issue more rigorously in Sect. \ref{XXZ}, where injection of the bodies from the inner OC (in the classical jargon) is studied. Since MD enhances binding energy of a comet, the classical effect of the Jupiter-Saturn barrier, in fact, has to be revised, see Sect. \ref{JSBinMOND}. Last but not least we have to emphasise that the steady-state distribution of the bodies in the cloud could look different in MD, see discussion in Sect. \ref{sum}.

\begin{figure}\centering
\resizebox{0.55\hsize}{!}{\includegraphics{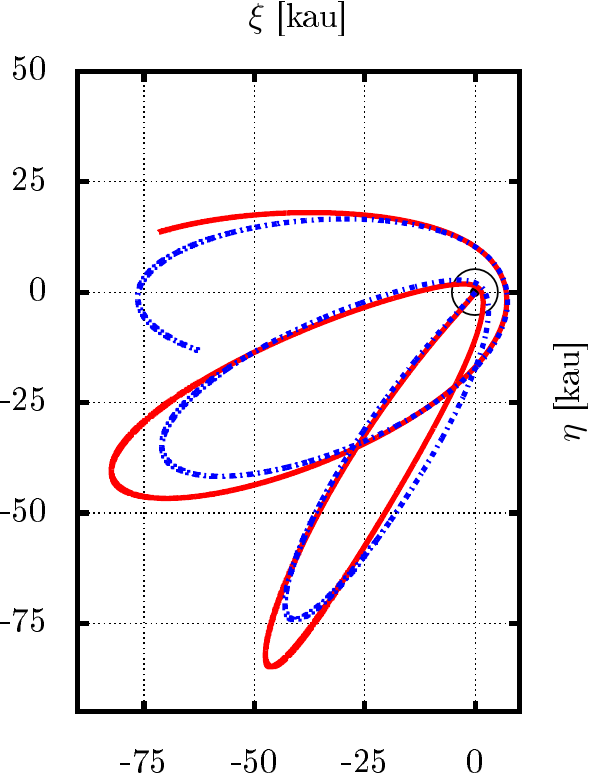}}
\caption{Past trajectories of two slightly hyperbolic comets in the simple model of the MOC. Both were initialised at their perihelia, one with $q=8$ au, $e=1.00150$, $\omega=\pi/4$ (solid line), the other with $q=3$ au, $e=1.00055$, $\omega=\pi/4$ (dot-dashed line). All the other orbital elements were set to 0. The integration time was 20 Myr. As can be seen these comets are bound (returning) in MD. The Sun resides at [0,0], as indicated by the symbol.}
\label{fig:fig2}
\end{figure}

\subsection{Escaping comets?}\label{escapingcomets}
We use the term ``hyperbolic comet'' for a comet whose Newtonian two-body orbital energy is positive and which is, according to a Newtonian astronomer, not bound (not returning) to the solar system. In this section, we investigate the idea that slightly hyperbolic comets can be bound to the Milgromian solar system, as first pointed out by \citet{Mil86b}.

The statistics of the original reciprocal semi-major axes, $1/a_{orig}$, also reveals, besides the famous Oort spike, a small but non-negligible number of slightly hyperbolic comets ($e$ slightly larger than 1; e.g. Fig. 1b in \citet{Don+04}). These are usually considered to follow very eccentric elliptic orbits in reality, rather than to be interstellar intrudes, but owing to observational errors or the inappropriate modelling of non-gravitational forces, they seem to occupy hyperbolic orbits \citep{Don+04}. Thanks to the boosted gravity in MD, the slightly hyperbolic comet could be still bound to the solar system\footnote{To be thorough, in Newtonian dynamics it is vice-versa possible for a comet to appear to be bound but to originate in the interstellar space as a result of Galactic tidal influence \citep{NJ13}. In any case, this special configuration is highly improbable \citep{NJ13}.}.

Comparing the escape speed at perihelion, $V_{esc}(q)$, see Eqs. (\ref{simplealgebra}) and (\ref{vesc}), with the tangential speed at the perihelion, $V_{peri}(e,q)$, we can decide whether a comet is bound or not. $V_{peri}(e,q)$ can be computed in the usual way. We are at the perihelion - in the deep Newtonian regime, and it depends only on the local gravitational field.
The opposite case is the escape speed, which has to be calculated from the MD gravity, no matter where we start from, see Eq. (\ref{vesc}). Assuming motion in the ecliptic plane, $i=0$, we have
\begin{eqnarray}\label{Cec}
V_{peri}~=~\sqrt{\frac{GM_{\odot}}{q}\left(1+e\right)}~.
\end{eqnarray}
Radial speed at the perihelion is 0. Thus for a given $q$, we can find the limiting eccentricity $e_{lim}$, so $e>e_{lim}$ implies $V_{peri}(e,q)>V_{esc}(q)$ . For example $q=3~au$ implies $e_{lim}=1.00075$ and $q=8~au$ leads to $e_{lim}=1.00199$. Slightly hyperbolic comets with $e<e_{lim}$ are bound in MD. Fig. \ref{fig:fig2} shows the trajectories of two comets that were initialised with the orbital elements $q=3$ au, $e=1.00055$,  $\omega=\pi/4$ (all the other elements are set to 0) and $q=8$ au, $e=1.00150$, $\omega=\pi/4$ (all the other elements are set to 0), and then integrated backwards for 20 Myr, assuming the simple model of the MOC. This is quite a long time interval to assume  the stationarity of the external field, thus the real trajectories would be a little different, as the external field changes its direction. In any case, we only intend to illustrate as slightly hyperbolic comets can be bound in MD, and this qualitative result remains the same. 

Observations of comets with similar original orbital elements could inflate the former conservative estimate of the MOC size to sizes comparable with the classical OC.
In Sect. \ref{simul} we take real cometary data and look at what they say about the size and shape of the MOC.

\subsection{Do Jupiter and Saturn act as a barrier in MD?}\label{JSBinMOND}
The enhanced binding energy of MOC comets raises a question: how does the mechanism of the planetary barrier that is operating in the classical OC change in the MD case?

QUMOND conserves energy. We use Eqs. (\ref{simplealgebra}) and (\ref{vesc}) to approximate QUMOND and assume energy conservation. We take a comet at perihelion, lying deeply in the Newtonian regime, with kinematics characterised by the Newtonian orbital elements, $a$ and $q$. We can find its specific binding energy in MD, simply as
\begin{eqnarray}\label{orb}
E_{BM}~=~-\frac{1}{2}\left[V_{peri}^{2}(a,q)-V^{2}_{esc}(q)\right],
\end{eqnarray}
where we can use Eq. (\ref{Cec}) under the assumption $i=0$. We note that we have put a minus sign in front of the factor 1/2 on the RHS of Eq. (\ref{orb}) because the binding energy is defined as a positive number.
For comets with $a=10$, 50, and 100 kau, the ratio $E_{BM}/E_{BN}$, where $E_{BN}=\left[GM_{\odot}/(2a)\right]$ is the Newtonian binding energy per unit mass, is approximately equal to 3, 13, and 26 respectively. Using the 1D QUMOND approximation, Eq. (60) in \citet{FM12}, instead of Eq. (\ref{simplealgebra}), these ratios are 2, 7, and 13 respectively. For near-parabolic orbits the value of $E_{BM}$ depends only weakly on $q$.

A comet of the classical OC in a typical orbit of, for example, $a=50$ kau, experiences an energy change per perihelion passage proportional to its own binding energy\footnote{This certainly depends on the orbital inclination, as can be seen in Fig. 1 in \citet{FB00}. The footnoted sentence is true for highly inclined orbits with $i\in(120,~150)~\deg$. For orbits close to ecliptics, the planetary kick at 15 au is about 6 times larger.} at $q\sim15$ au, see Fig. 1 in \citet{FB00}. Making the binding energy of this comet in MD $\sim10$ times larger this criterion is met at $q\sim7$ au. Roughly speaking this means that MOC comets with $q<7$ au, instead of the classical value  $\sim15$ au, are removed from the cloud due to planetary perturbations. The planetary barrier similarly to the whole cloud shifts inward in MD. Anyway it can still act in a way of inflating semi-major axes for those comets having $q>7$ au, but these are not a priori prevented from being injected inside the inner solar system as in the case of the removed comets of the classical OC.

\begin{figure*}
\begin{center}
\resizebox{0.75\hsize}{!}{\includegraphics{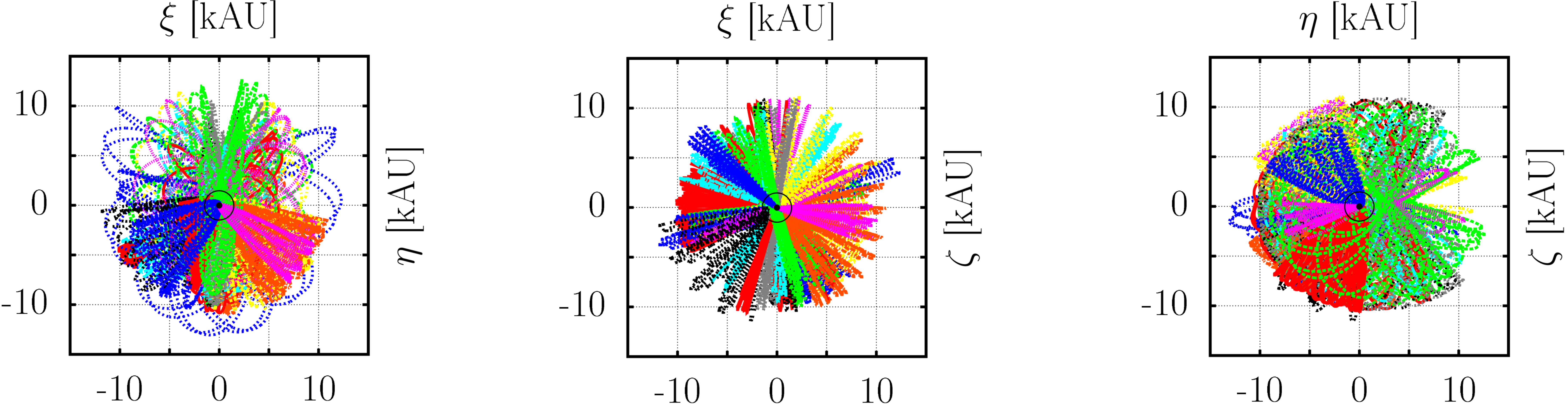}}
\caption{Past Milgromian trajectories of 31 near-parabolic comets, those identified as dynamically new in \citetalias{DK11}, projected to 3 mutually orthogonal planes of $O_{\odot}(\xi,\eta,\zeta)$.
Dynamical model of the OC includes the stationary Galactic field coupled to the QUMOND equations, see Sect. \ref{sec:simpleOC}, and the Galactic tide model, see Sect. \ref{tide}. 
The comets with Keplerian periods $T_{Kep}$ lesser than 10 Myr were followed for the time of $T_{Kep}$, those with $T_{Kep}>$ 10 Myr were followed for 10 Myr. Inferred MOC is much smaller than the classical OC, see Table \ref{comets} for comparison with Newtonian orbits. At [0,0] resides the Sun as indicated by the symbol.}
\label{img:nearp}
\end{center}
\end{figure*}
\begin{figure*}
\begin{center}
\resizebox{0.75\hsize}{!}{\includegraphics{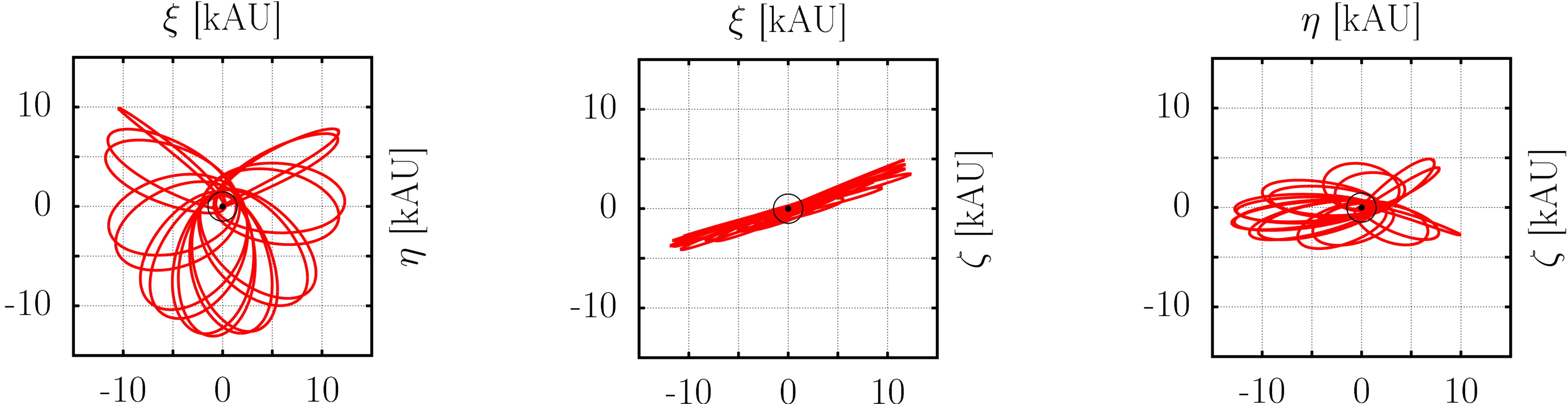}}
\caption{Past Milgromian trajectory of the comet C/1978 G2, the slightly hyperbolic comet. Initial $q=6.28$ au and $e$ = 1.00014083. At [0,0] resides the Sun as indicated by the symbol.}
\label{img:1978g2}
\end{center}
\end{figure*}

\section{Observed near-parabolic comets in Milgromian dynamics}\label{simul}
Motivated by the crude picture of the OC outlined in Sect. \ref{MilgromianOC}, we have used real cometary data to investigate origin of the near-parabolic comets in the framework of MD.

We have approximated action of QUMOND by the simple model of the MOC, with the constant external field of the Galaxy ${\bf g_{e}}$ coupled to the QUMOND equations, see Sect.~\ref{sec:simpleOC}. The rotation of ${\bf g_{e}}$ has period of $\sim210$ Myr, therefore we use integration times to be Keplerian periods for those comets having these lesser than 10 Myr. For those that have Keplerian periods larger than 10 Myr we use integration time of 10 Myr as all these have much shorter real (QUMOND) ``periods'', i.e. times between two successive perihelia.
Moreover, the tidal effect, which comes from the Galactic gravity gradient across the OC, is also accounted. The Galactic tide model is described in Sect. \ref{tide}. This model reflects the local density of the baryonic + phantom matter as determined by QUMOND for the adopted baryonic model of the Galaxy, see sections \ref{PMD} and \ref{Galaxymodel}. 
We have simply added the tidal acceleration (0, 0, $\ddot{\zeta}_{tide}$), Eq. (\ref{tidesimple}), to RHS of Eq. (\ref{eqmo}). This is only an approximation in non-linear MD. 
But, it proves to be good idea to model EFE (assuming a spatially invariant field) and tides as two separate effects of the Galaxy, see sections \ref{results} and \ref{amch}.

We have taken the original orbits from the sample of near-parabolic comets that were identified as dynamically new in \citetalias{DK11}, converted them to initial positions and velocities of test particles, and integrated these back in time, looking for their past Milgromian trajectories.

\subsection{Data}\label{data}
Our sample consists of the 31 comets identified as dynamically new in \citetalias{DK11}. We omitted errors in the lengths of original semi-major axes $a_{orig}$, the only orbital element with significant error and, instead, only took their expected values as these are fairly typical for Oort spike comets. A more exact approach would proceed in a similar manner as \citetalias{DK11} did, covering the error in the orbital energy determination with a large number of virtual orbits, but this is much more processor-time consuming in MD than in Newtonian dynamics.

The sample also contains one slightly hyperbolic comet, C/1978 G2, with perihelion $q=6.28$ au and eccentricity $e= 1.00014083$. We also note the orbit of the comet C/2005 B1, with a very large semi-major axis of 250.6 kau.

Original orbital elements of sample comets were retrieved from \citet{Kro14} and are displayed in Table \ref{comets}. These were calculated at a heliocentric distance 250 au, which is still well within the Newtonian regime.

\subsection{Results}\label{results}
The past QUMOND trajectories of the sample comets are shown in Fig. \ref{img:nearp}. The resulting size and overall shape of the MOC is in large agreement with the one obtained in Sect. \ref{MilgromianOC}. The trajectory of the single comet with $e>1$ in our sample, C/1978 G2, is redrawn in Fig. \ref{img:1978g2}. In Milgromian framework the comet is bound, visiting similar heliocentric distances to the other comets in our sample.

In MD, we expect the Galactic tide to be stronger than in the Newtonian dynamics, see Fig. \ref{img:pmd} and Sect. \ref{tide}.
However, the changes in orbits - perihelia positions and precession rates - induced by the Galactic tide are negligible, compared to those induced by the EFE, see also Sect. \ref{XXZ}. Figs. \ref{img:nearp} and \ref{img:1978g2} would not look different if the Galactic tide model as presented in Sect. \ref{tide} was not incorporated. This is a natural consequence of the compactness of the cloud. The comets cruise up to $\Xi\sim$ 13 kau, where the tidal torquing is still minute, but EFE plays a dominant role. As mentioned above, we model the EFE and the Galactic tide as the separate effects. 

\begin{figure}
\begin{center}
\resizebox{0.65\hsize}{!}{\includegraphics{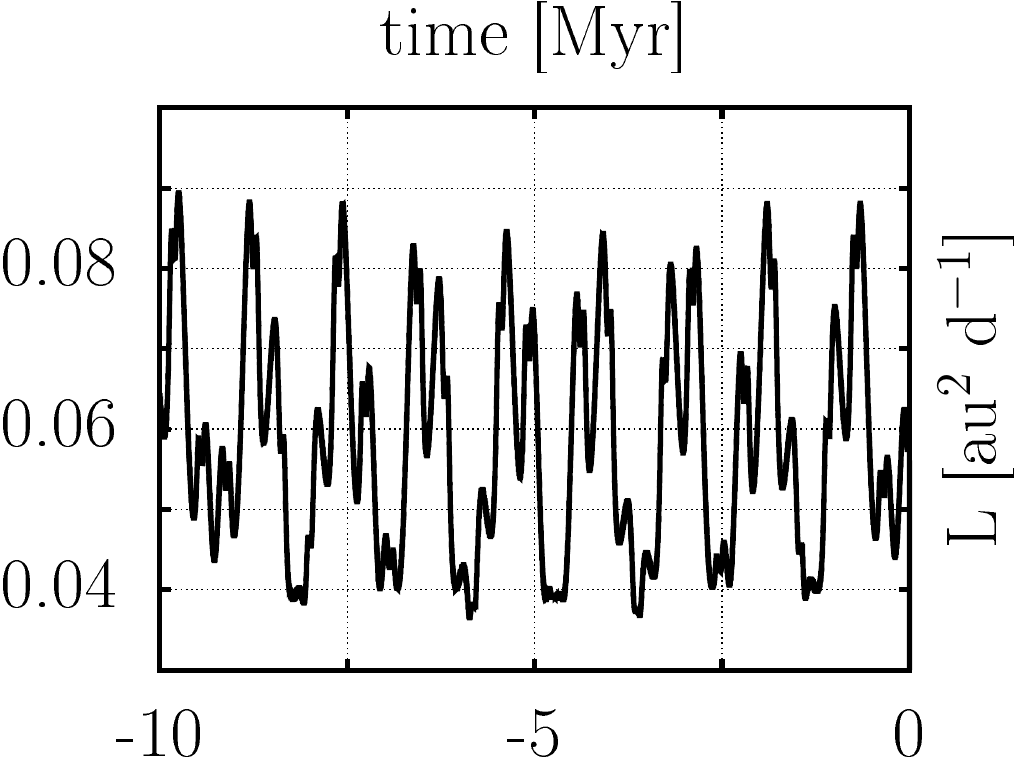}}
\caption{Specific angular momentum $L$ as a function of time for the comet C/1974 V1. We have assumed the simple model of the MOC (tides are omitted). The periodic changes are induced solely by EFE. The negative time means that we are dealing with the past trajectory of the comet.}
\label{img:L1974v1}
\end{center}
\end{figure}

In Fig. \ref{img:L1974v1} we show the specific angular momentum as a function of time, $L(t)$, for the comet C/1974 V1 in the simple model of the MOC. Tides are omitted this time. Periodic changes in angular momentum are induced purely by EFE. Similar behaviour can also be found by checking the other comets in the sample. Taking into account the Galactic tide only has a minor effect and $L(t)$ is very much the same.

\section{Galactic torque}\label{XXZ}
We have shown that MOC is much smaller than the classical OC. The MOC boundary, as found by tracing Oort spike comets with an initial eccentricity $e<1$ (which is the vast majority of observed comets) back in time, lies at heliocentric distances that correspond to the classical inner OC. Also the single comet with $e>1$ in the Sect. \ref{simul} sample, C/1978 G2, orbits in bound orbit at similarly small heliocentric distances in MD. It is presumed that the tidal force at these heliocentric distances is not large enough to sufficiently quickly decrease perihelion distance so that a comet bypasses the Jupiter-Saturn barrier, e.g. \citet{Don+04}. In MD, the compactness of the OC does not need to be an obstacle for the injection of a comet into the inner solar system because of the action of EFE. 

\subsection{Angular momentum change}\label{amch}
In this section, we preserve the classical idea of the Jupiter-Saturn barrier at $\sim$ 15 au, although in Sect. \ref{JSBinMOND} we have shown that the barrier actually shifts inwards in MD. This shift naturally increases the inflow of comets.

To illustrate the capability of the EFE to deliver OC bodies into the inner solar system, we have run similar simulation to those in Sect. \ref{MilgromianOC}. In this case, we intended to mimic the sample of comets that are about to enter/leave the planetary zone. Consequently, we chose the initial perihelion distance of each particle, $q$, to be a random number that is uniformly distributed on the interval $(15,100)$ au. All the other initial orbital elements of the test particles were randomly generated in the same way as in Sect. \ref{MilgromianOC}.
The orbital elements were at $t=0$, transformed to initial cartesian positions and velocities, the real observables. 

We employed two distinct dynamical models of the OC, one of which is Milgromian and the other, Newtonian: (i) the simple model of the MOC, and, (ii) Sun + Galactic tide in the Newtonian framework.
We tested the fact that incorporation of the Galactic tide model, as described in Sec. \ref{tide}, into the simple model of the MOC has negligible effects for the times that correspond to one revolution of a comet. This is obviously because the comets of the MOC orbit in $\Xi\lesssim$ 15 kau, at these heliocentric distances, the tidal force is too weak. Two distinct $\varrho(z_{\odot})$ were used: $\varrho(z_{\odot})=\varrho_{b}(z_{\odot})+\varrho_{ph}(z_{\odot})$ in MD and $\varrho(z_{\odot})=\varrho_{b}(z_{\odot})+\varrho_{h}(z_{\odot})$ in Newtonian dynamics, where $\varrho_{b}(z_{\odot})$ is the local vertical density of baryons and $\varrho_{h}(z_{\odot})$ is the local vertical density of the NFW DM halo.
 
\begin{figure}
\begin{center}
\resizebox{\hsize}{!}{\includegraphics{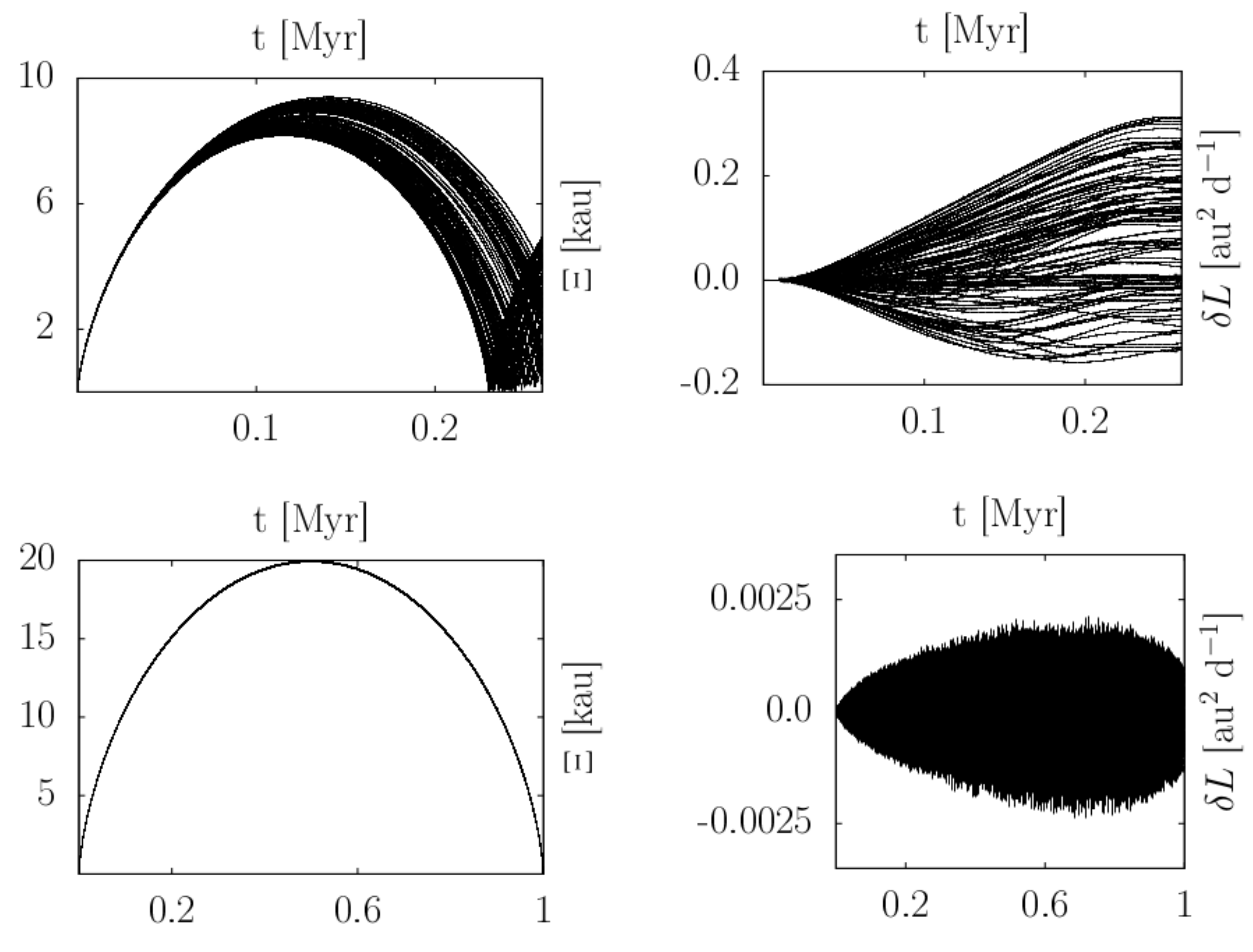}}
\caption{Heliocentric distance, $\Xi(t)$, and change in magnitude of the specific angular momentum, $\delta L(t)\equiv L(t)-L(0)$, as a function of time, $t$, for 100 Monte Carlo test particles initialised with $a=10$ kau, and $q$ uniformly distributed on the interval $(15,100)$ au. The top row represents an output of the Milgromian simulation, the bottom row, the Newtonian simulation. In MD simulation, the follow up time, $T_{rev}$, is set to 0.26 Myr (see top left quarter of the figure for motivation), in Newtonian simulation $T_{rev}$ is set to be the Keplerian period $T_{Kep}$($a$=10 kau) $\approx$ 1 Myr.}
\label{img:dl00}
\end{center}
\end{figure}

\begin{figure*}
\begin{center}
\resizebox{0.65\hsize}{!}{\includegraphics{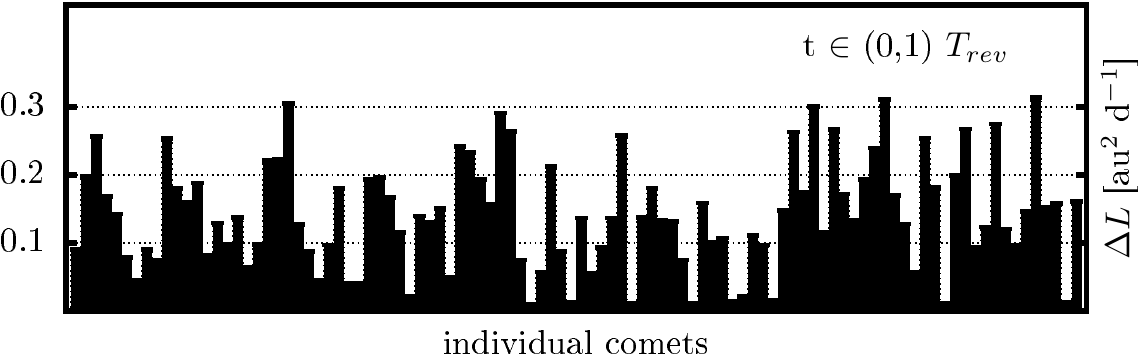}}
\caption{Histogram of $\Delta L\equiv L_{max}-L_{min}$ for 100 Monte Carlo test particles initialised with $a=$10 kau and $q$ uniformly distributed on the interval $(15,100)$ au. Here $L_{max}$ ($L_{min}$) is maximal (minimal) magnitude of the specific angular momentum, as found during one revolution, $T_{rev}$, succeeding the initialisation of a comet at perihelion. In the MD simulation $T_{rev}=0.26$ Myr, in the Newtonian simulation $T_{rev}=T_{Kep}$($a$ = 10 kau) $\approx$ 1 Myr. A single bin corresponds to a single test particle in the simulation. Solid bins are $\Delta L$ in the simple model of the MOC, shaded bins (here barely visible), stacked on the solid bins, are $\Delta L$ in Newtonian dynamics, with the gravity of the Sun and the Galactic tide accounted for.}
\label{img:dl0}
\end{center}
\end{figure*}

\begin{figure}
\begin{center}
\resizebox{\hsize}{!}{\includegraphics{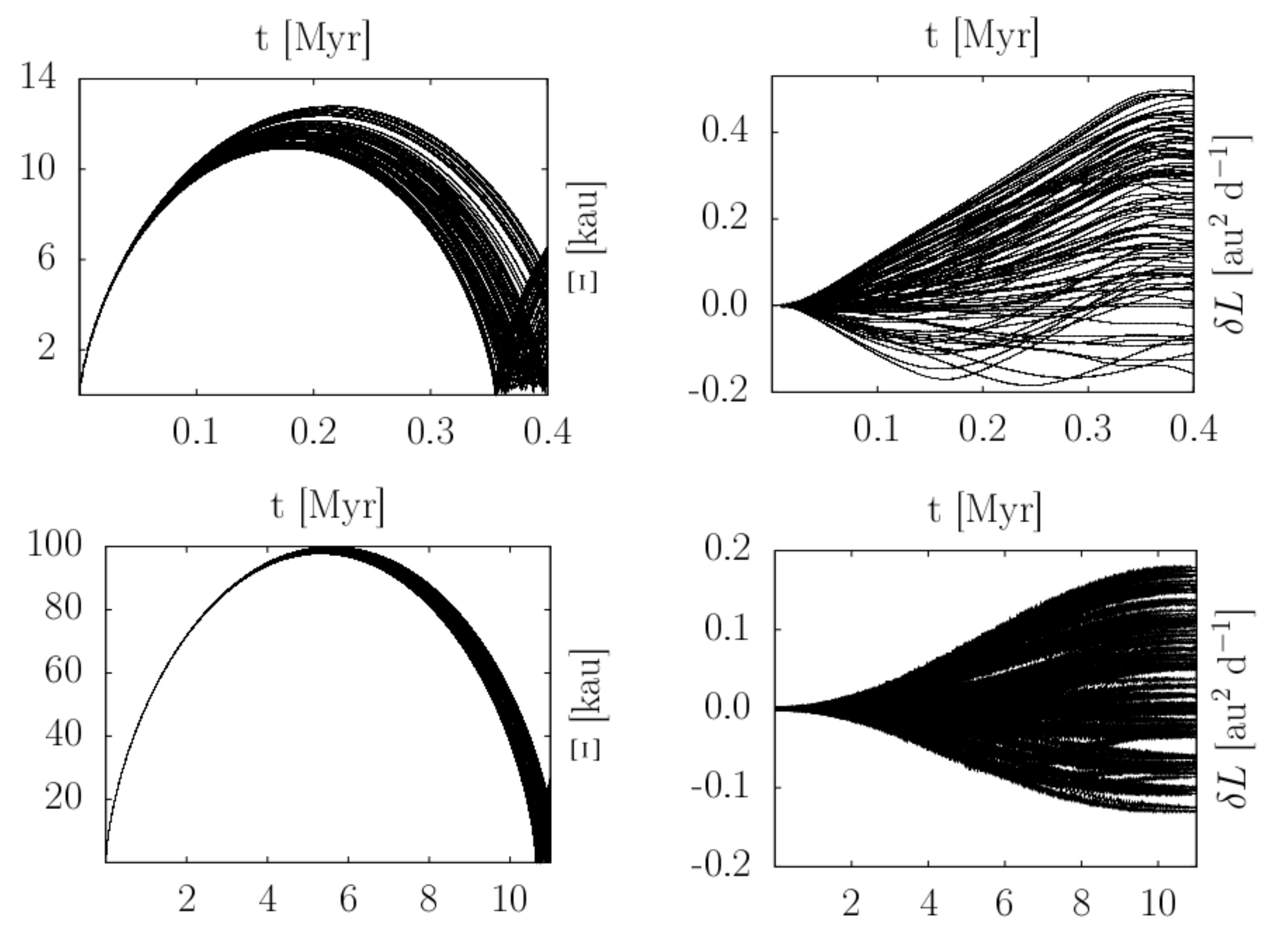}}
\caption{Same as for Fig. \ref{img:dl00}, but now the particles are initialised with $a=50$ kau. The top row represents an output of the Milgromian simulation, the bottom row, the Newtonian simulation. In MD simulation, the follow-up time, $T_{rev}$, is set to 0.4 Myr (see top left quarter of this figure for motivation), in the Newtonian simulation $T_{rev}$ is set to be the Keplerian period $T_{Kep}$($a$=50 kau) $\approx$ 11 Myr.}
\label{img:dl11}
\end{center}
\end{figure}

\begin{figure*}
\begin{center}
\resizebox{0.65\hsize}{!}{\includegraphics{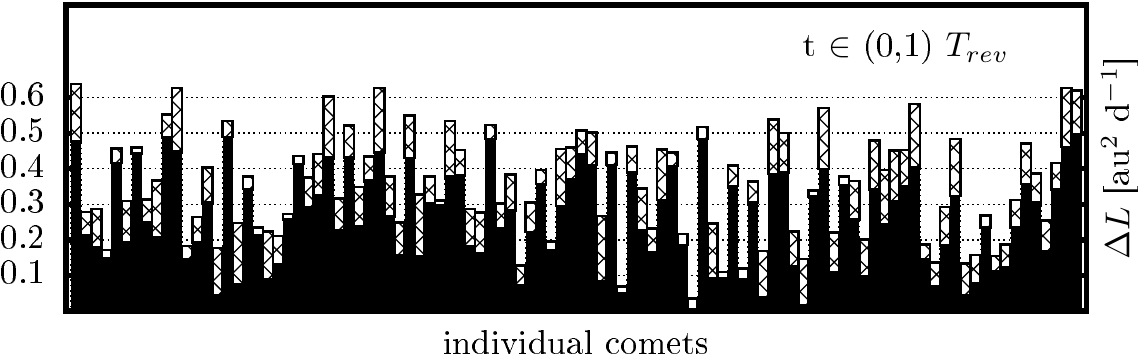}}
\caption{Same as for Fig. \ref{img:dl0} but now the particles are initialized with $a=50$ kau. In the MD simulation $T_{rev}=0.4$ Myr, in the Newtonian simulation $T_{rev}=T_{Kep}$($a$ = 50 kau) $\approx$ 11 Myr. A single bin corresponds to a single test particle in the simulation. Solid bins are $\Delta L$ in the simple model of the MOC, shaded bins, stacked on the solid bins, are $\Delta L$ in Newtonian dynamics, with the gravity of the Sun and the Galactic tide accounted for.}
\label{img:dl1}
\end{center}
\end{figure*}

\begin{figure}
\begin{center}
\resizebox{\hsize}{!}{\includegraphics{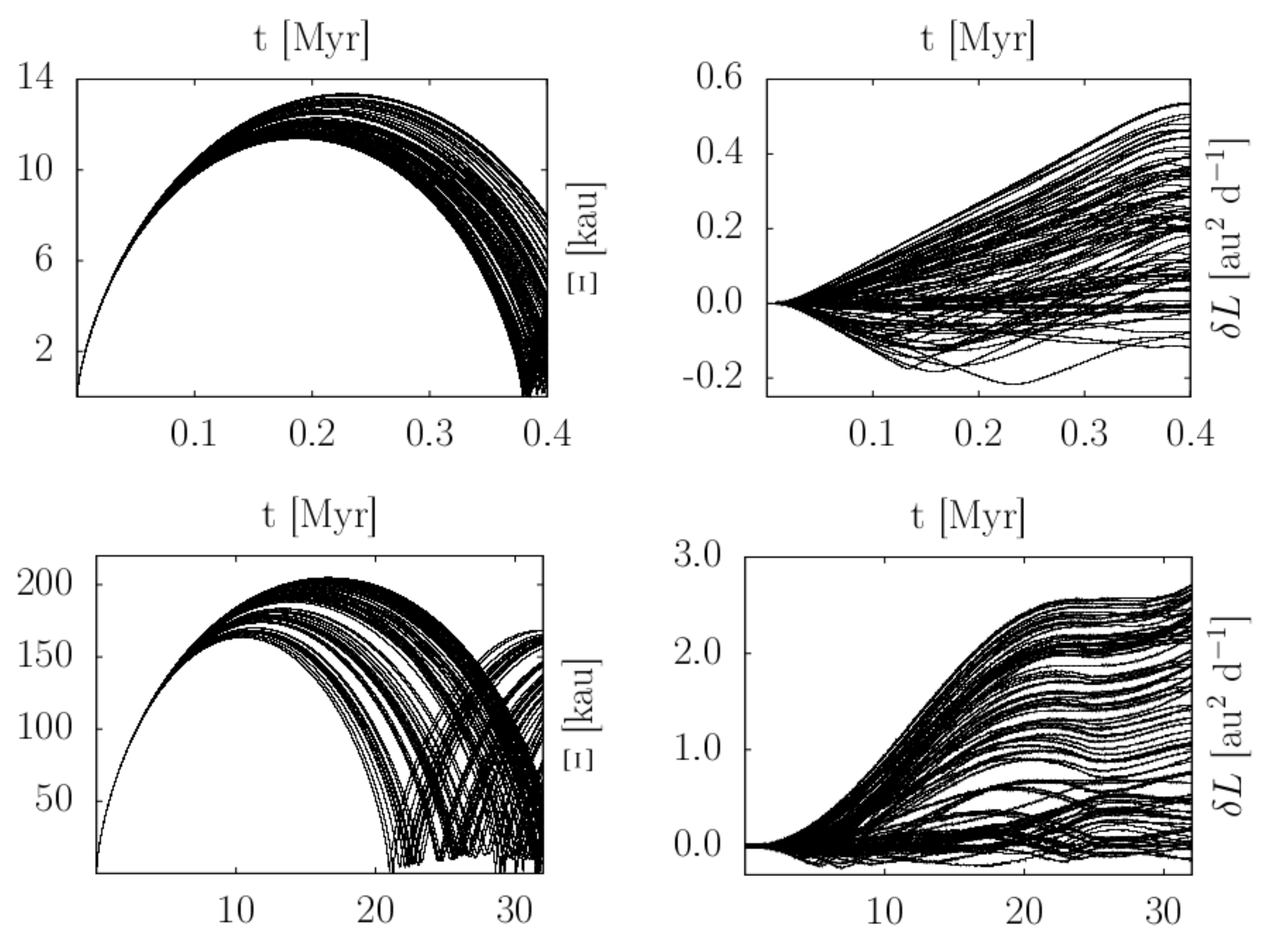}}
\caption{Same as for Fig. \ref{img:dl00}, but now the particles are initialised with $a=100$ kau. The top row represents an output of the Milgromian simulation, the bottom row, the Newtonian simulation. In the MD simulation, the follow-up time, $T_{rev}$, is set to 0.4 Myr (see top left quarter of this figure for motivation), in the Newtonian simulation, $T_{rev}$ is set to be the Keplerian period $T_{Kep}$($a$=100 kau) $\approx$ 32 Myr.}
\label{img:dl22}
\end{center}
\end{figure}

\begin{figure*}
\begin{center}
\resizebox{0.65\hsize}{!}{\includegraphics{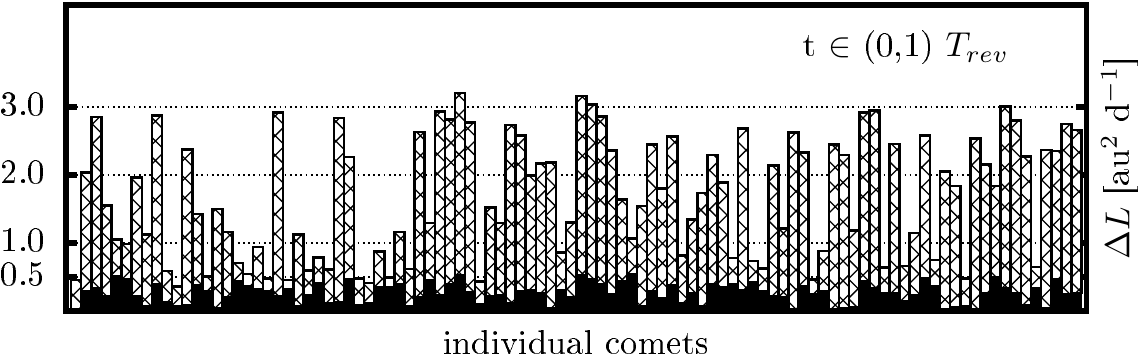}}
\caption{Same as for Fig. \ref{img:dl0}, but now the particles are initialized with $a=100$ kau. In the MD simulation $T_{rev}=0.4$ Myr, in the Newtonian simulation, $T_{rev}=T_{Kep}$($a$ = 100 kau) $\approx$ 32 Myr. A single bin corresponds to a single test particle in the simulation. Solid bins are $\Delta L$ in the simple model of the MOC, shaded bins, stacked on the solid bins, are $\Delta L$ in Newtonian dynamics, with the gravity of the Sun and the Galactic tide accounted for.}
\label{img:dl2}
\end{center}
\end{figure*}

Figs. \ref{img:dl00} ($a=10$ kau), \ref{img:dl11} ($a=50$ kau), and \ref{img:dl22} ($a=100$ kau) show the heliocentric distance, $\Xi(t)$, and change in magnitude of the specific angular momentum, $\delta L(t)\equiv L(t)-L(0)$,  of the particles, as a function of time. The followed time window, $T_{rev}$, corresponds approximately to one revolution that succeeds the perihelion initialisation. In Figs. \ref{img:dl0} ($a=10$ kau), \ref{img:dl1} ($a=50$ kau), and \ref{img:dl2} ($a=100$ kau) we show the value of $\Delta L\equiv L_{max}-L_{min}$ of the individual particles, where $L_{max}\equiv [L(t)]_{max}$ and $L_{min}\equiv [L(t)]_{min}$ are the maximal and the minimal value of $L(t)$ during $T_{rev}$.

When interpreting these figures, we have to bear in mind the timescales of the angular momentum changes, these are $\sim4$ ($a=10$ kau) to $\sim80$ ($a=100$ kau) times smaller in the MOC than in the classical OC. We also note that the particles that are initialised with $a$ as large as 100 kau are travelling in $\Xi\lesssim 15$ kau in the MOC.
It is evident that the injection could be very efficient in the MOC, nevertheless the MOC is much more radially compact than the classical OC. In MD, the rapid changes in the angular momentum are induced by EFE. Moreover the bodies that are hidden in the classical OC - i.e. not able to reach the observability region, because of either their immunity  from the action of the external perturbers, the hypothesized inner core, or, the inability to overshoot the Jupiter-Saturn barrier, the inner OC bodies with $a\sim 10$ kau, can, because of EFE, also be delivered from the MOC into the inner solar system, see also Sect. \ref{Sedna}. 

Figs. \ref{img:dl1} and \ref{img:dl2} show that Newtonian tides (OC) overcome EFE (MOC) in $\Delta q$ per revolution only for comets with $a$ as large as $\sim$ 50 - 100 kau. This is 9 out of 30 comets with $e<1$ in the Sect. \ref{simul} sample. 

\subsection{Sedna}\label{Sedna}
We have shown that cometary perihelia can be very effectively torqued in and out by EFE, even for those comets that are travelling in fairly small heliocentric distances, $\sim10$ kau.
Is torquing due to EFE important at even smaller heliocentric distances? Is EFE responsible for the shape of the current puzzling orbit of the trans-Neptunian planetoid Sedna? To address these questions we ran the following simulation: 100 Monte Carlo test particles (Sedna progenitors - Sednitos) with initial $a=524$ au (Sedna's heliocentric $a$ at epoch 2,457,000.5 JD, according to JPL's service {\small HORIZONS}) and, among the particles, uniformly distributed $q$ in bounds $(5,30)$ au, $i$ in bounds $(0,10)~\deg$, $\omega$ and $\Omega$ in bounds $(0,2\pi)$, were initialised at their perihelia and then followed for 5.9 Myr in the simple model of the MOC. These initial orbital elements have been chosen to mimic the protoplanetary disk origin of Sedna. We assumed that Sednito's semi-major axis was already pumped to the current Sedna's value at the beginning of the simulation, owing to past planetary encounters. The planets were omitted in the simulation.

\begin{figure}\centering
\resizebox{0.7\hsize}{!}{\includegraphics{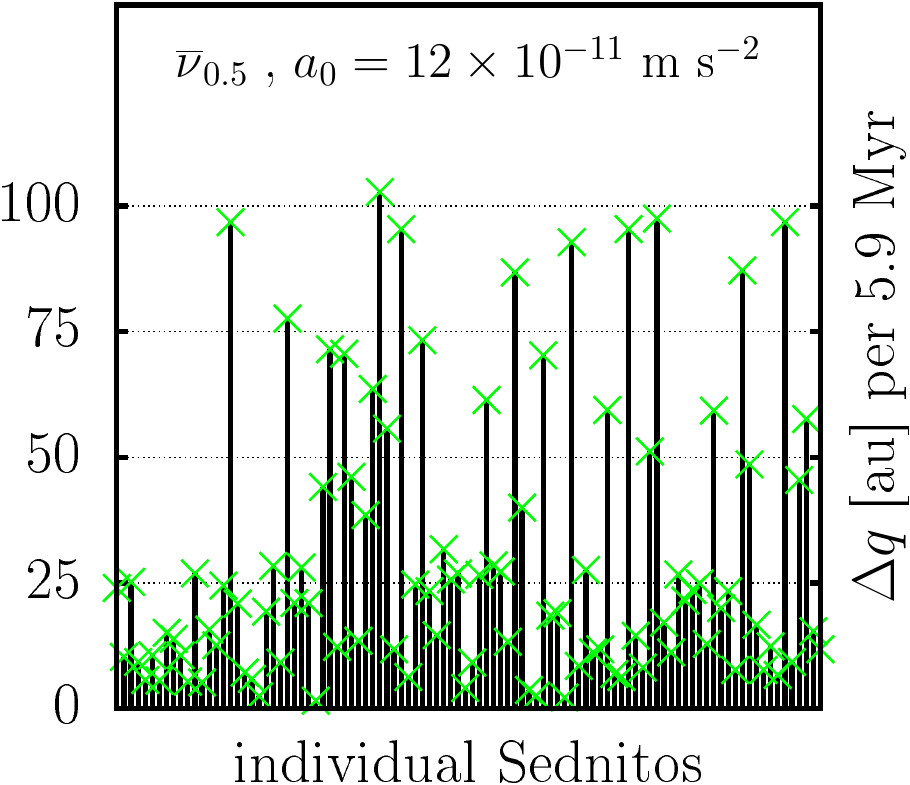}}
\caption{$\Delta q \equiv q_{max}-q_{min}$ per 5.9 Myr for 100 Sedna progenitors (Sednitos) moving under the action of EFE. Sednitos were initialised at perihelia with $a=524$ au, uniformly distributed $q\in(5,30)$ au, $i\in(-10,10)\deg$, $\omega$ and $\Omega\in(0,2\pi)$.}
\label{img:Sedna}
\end{figure}

In Fig. \ref{img:Sedna} we show $\Delta q \equiv q_{max}-q_{min}$ for 100 simulated Sednitos, where $q_{max}$ and $q_{min}$ are Sednito's maximal and minimal value of $q$, per 5.9 Myr. As can be seen in some cases, $\Delta q$ is as large as $100$ au. At the end of the simulation, 7 Sednitos had $q\sim75$ au, hence very close to the Sedna's perihelion distance. Sedna-like orbits (here simplified as specific $a$ and $q$ values) can be produced by EFE in a few Myr. The catch is that, as $q$ oscillates in and out on timescales of millions of years, the trans-Neptunian bodies with similar orbits as Sedna could possibly wander into the inner solar system. It is also possible that substantial migrants have already been removed from these orbits and the current population is relatively stable against migration.

\begin{figure}
\begin{center}
\resizebox{0.7\hsize}{!}{\includegraphics{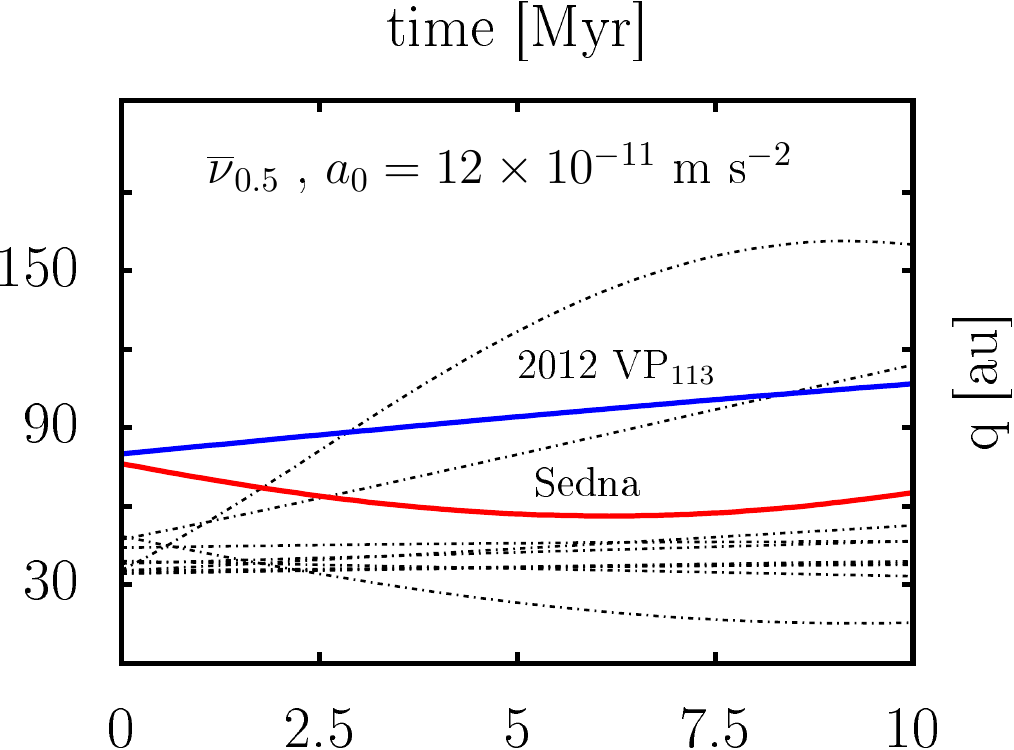}}
\caption{Perihelion distance as a function of time under the action of EFE for the known population of trans-Neptunian bodies with $a>150$ au and $q>30$ au. Thick solid lines are 2012 VP$_{113}$ and Sedna.}
\label{img:Sednitos}
\end{center}
\end{figure}

Fig. \ref{img:Sednitos} depicts perihelion distance as a function of time for all known bodies with $q>30$ au and $a>150$ au in the simple model of the MOC during the next 10 Myr. Initial orbital elements of the bodies were retrieved from \citet{TS14}, see their Table 1 and extended data Table 2. Only one of the followed objects, 2010 GB$_{174}$, migrates under 30 au in the next 10 Myr. To investigate the migration of these objects thoroughly, we would have to improve our dynamical model of the MOC to account for the change in the external field direction, since long integration times would be necessary, and also to account for the planetary perturbations. We leave this task to our future studies.


\section{Varying interpolating function and $a_{0}$}\label{if}
\citet{Hee+15} (hereafter \citetalias{Hee+15}) recently constrained the most frequently used families of the MD interpolating (transition) function (e.g. Sect. 6.2 in \citealp{FM12}) with the Cassini spacecraft radio tracking data \citep{Hee+14}. These constraints come from EFE, which produces small quadrupole correction to the Newtonian potential in the planetary region.
They concluded the following constraints (on $n$):
\begin{subequations}
\begin{align}
        \nu_{n}(\beta) 
&~=~\left[\frac{1+\left(1+4\beta^{-n}\right)^{1/2}}{2}\right]^{1/n}~~~~~~~~~~~~~,~~~n\geq7~,\label{if1}\\
        \widehat{\nu}_{n}(\beta) &~=~\left[1-\exp(-\beta^{~n/2})\right]^{-1/n}~~~~~~~~~~~~~~~,~~~n\geq6~,\label{if2}\\
        \overline{\nu}_{n}(\beta) 
&~=~\left[\widehat{\nu}_{2n}(\beta)\right] +\left(1-\frac{1}{2n}\right)\exp(-\beta^{n})~~~,~~~n\geq2~,\label{if3}
\end{align}
\end{subequations}
where Eqs. (\ref{if1}) - (\ref{if3}) are three different families of the interpolating function $\nu$.
We note that $\widehat{\nu}_{1}=\overline{\nu}_{1/2}$. So far we have only used $\overline{\nu}_{1/2}$ in our calculations.
But according to the findings of \citetalias{Hee+15} this interpolating function is ruled out in the planetary region.. 
\citetalias{Hee+15} also revised the value of $a_{0}$, based on rotation curve fits (taking care whether EFE plays a role) and found optimum (best-fit) value for a given interpolating function. For example $\overline{\nu}_{n\geq 2}$ yields $a_{0}\lesssim 8.1\times 10^{-11}$ m s$^{-2}$, where the boundary value is a bit smaller than the standard value $a_{0}=1.2\times 10^{-10}$ m s$^{-2}$, but still well compatible with the baryonic Tully-Fisher or Faber-Jackson relation. In what follows, we consider only the $\overline{\nu}_{n}$ family.

\begin{figure}
\begin{center}
\resizebox{\hsize}{!}{\includegraphics{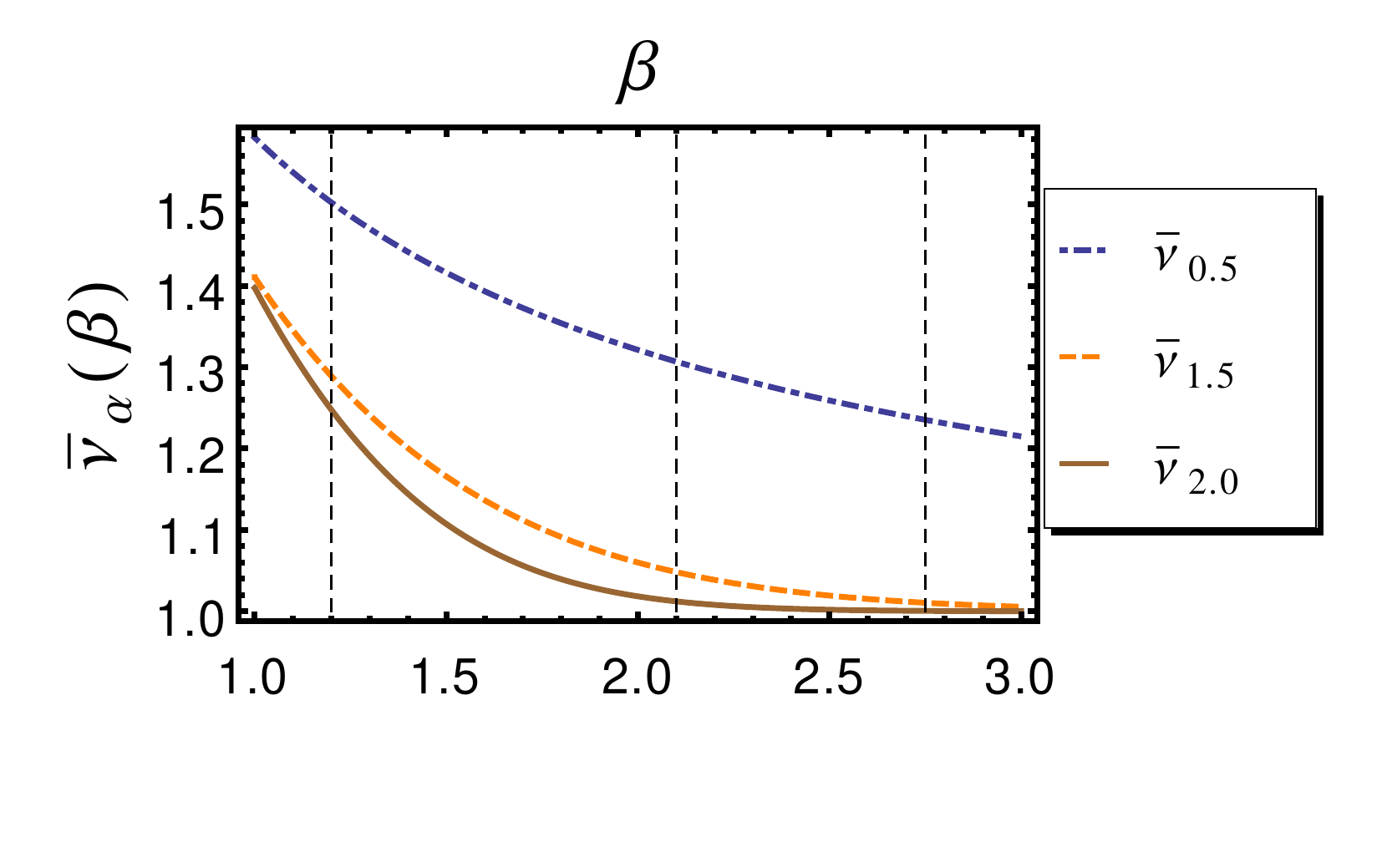}}
\caption{Transition to the Newtonian regime for different alphas in $\overline{\nu}_{\alpha}$ family. Vertical dashed lines, from left to right, mark values $\beta\equiv g^{N}/a_{0}=1.20$, 2.10 and 2.75, respectively.
These came from $g_{e}-\overline{\nu}_{\alpha}(g^{N}_{e}/a_{0})~g^{N}_{e}=0$, using $\alpha=0.5$,
$a_{0}=1.2\times10^{-10}$ m $s^{-2}$ ($\beta=1.20$);  $\alpha=0.5$,
$a_{0}=8.11\times10^{-11}$ m $s^{-2}$ ($\beta=2.10$); $\alpha=1.5$,
$a_{0}=8.11\times10^{-11}$ m $s^{-2}$ ($\beta=2.75$), and assuming that the external field dominates over the internal, $g^{N}\approx g^{N}_{e}$. When this is not the case the values of $\beta$ are even larger. Note that $\overline{\nu}_{1.5}(2.75)\approx\overline{\nu}_{2.0}(2.75)$.
$g_{e}$ is always the same constant $V^{2}_{0}/R_{0}$, but what matters is that the value of $g_{e}$ varies in units of $a_{0}$ as one varies $a_{0}$.}
\label{img:cassiniif}
\end{center}
\end{figure}

Fig. \ref{img:cassiniif} shows $\overline{\nu}_{\alpha}(\beta)$ for $1\leq\beta\leq3$ and three different alphas, $\alpha=0.5$, 1.5, and 2.0. The rightmost dashed vertical line indicates the smallest possible $\beta$ for the solar system in the field of the Galaxy assuming $\overline{\nu}_{1.5}(\beta)$ and $a_{0}=8.1\times10^{-11}$ m s$^{-2}$. This was found by assuming the external field dominance and solving for $g^{N}_{e}$ in $g_{e}-\overline{\nu}_{1.5}(g^{N}_{e}/a_{0})~g^{N}_{e}=0$, which yields $\beta=2.75$. If the internal field is non-negligible then $\beta$ is always larger. We note that $\overline{\nu}_{1.5}(2.75)\approx\overline{\nu}_{2.0}(2.75)$; as for this and the numerical convenience of using $\overline{\nu}_{1.5}$, we consider combination $\overline{\nu}_{1.5}$ and  $a_{0}=8.1\times10^{-11}$ m s$^{-2}$ in our calculations.

\begin{figure}
\begin{center}
\resizebox{0.85\hsize}{!}{\includegraphics{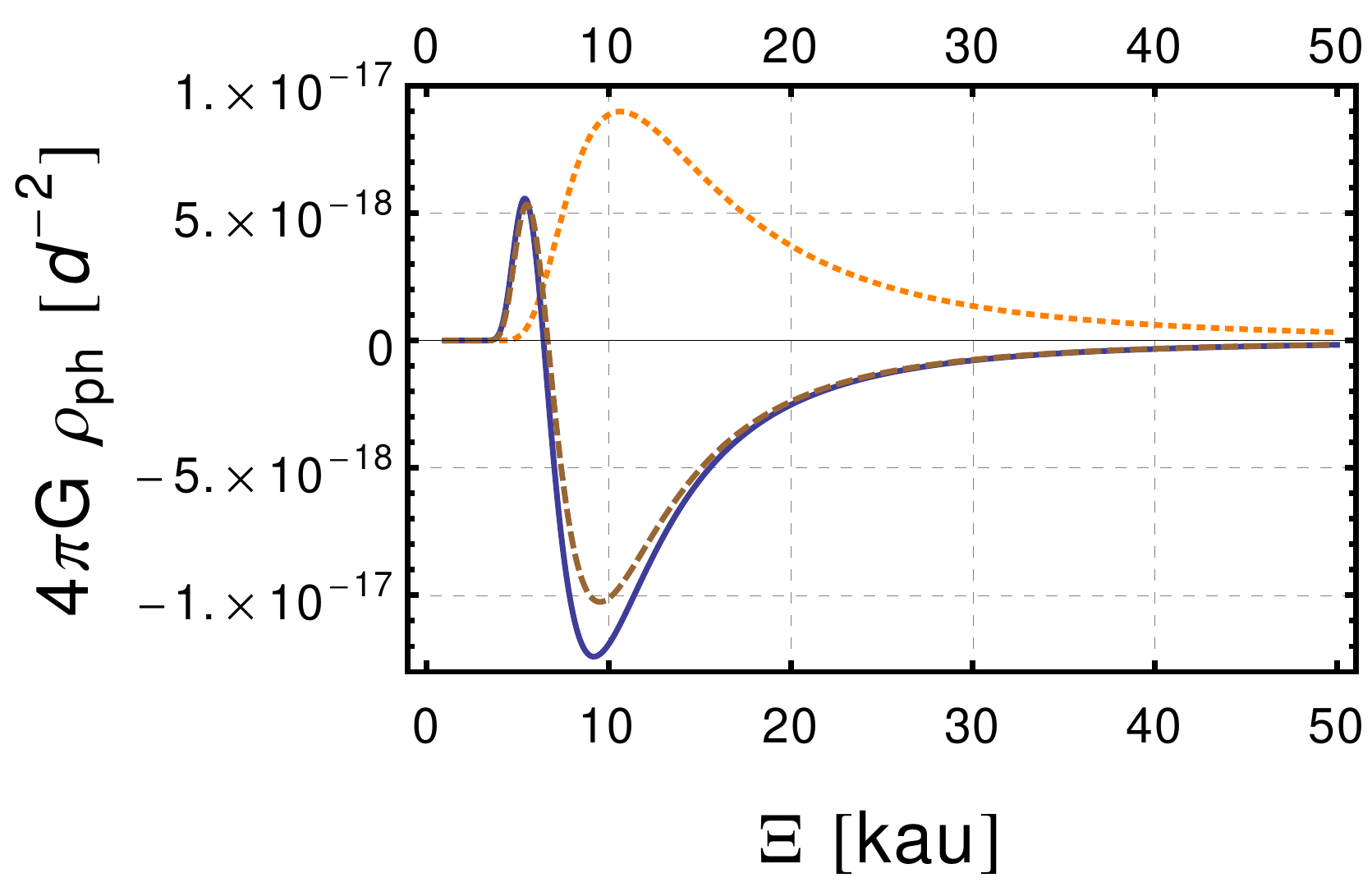}}
\caption{PMD in the simple model of the MOC in the direction of $\xi$ (solid line), $\eta$ (dotted line), and $\zeta$ (dashed line) axis of $O_{\odot}(\xi,\eta,\zeta)$. $\overline{\nu}_{1.5}$, and $a_{0}=8.11\times10^{-11}$ m $s^{-2}$ are assumed.}
\label{img:cassini2}
\end{center}
\end{figure}

PMD, Eq. (\ref{rophsimple}), as a function of heliocentric distance, $\Xi$, is depicted in Fig. \ref{img:cassini2} along $\xi$, $\eta$, and $\zeta$ axes of $O_{\odot}(\xi,\eta,\zeta)$. The simple model of the MOC, $\overline{\nu}_{1.5}(\beta)$ interpolating function and $a_{0}=8.1\times10^{-11}$ m s$^{-2}$, were assumed. The peaks in the positive values of $\varrho_{ph}$ are $\sim$800 (left) and $\sim$400 (right) times smaller than in the case of $\overline{\nu}_{0.5}(\beta)$ and $a_{0}=1.2\times10^{-10}$ m s$^{-2}$. We note that MD predicts the existence of regions with negative PMD in the solar system which is imposed to the gravitational field of the Galaxy \citep{Mil86a}.
This speciality of PMD makes MD possibly observationally distinguishable from the DM hypothesis.

We can conclude that, by adopting $\overline{\nu}_{\alpha}(\beta)$ with $\alpha=1.5$ or $2.0$ and  $a_{0}=8.1\times10^{-11}$ m s$^{-2}$ - or even larger $\alpha$ and smaller $a_{0}$ \citepalias{Hee+15}, we can expect the MOC to be very much Newtonian, since EFE, in this case, essentially suppresses the Milgromian regime at any distance from the Sun. The MOC is then of comparable overall size, comets have similar binding energies, and the Jupiter-Saturn barrier operates in a similar way to that found in Newtonian dynamics. Aphelia directions of observed, dynamically new comets were shown to avoid Galactic latitudes, $b_{G}$, close to the polar caps and the Galactic equator \citep{Del87}. This is conventionally considered to be a signature of the Galactic-tide-induced injection of the comets \citep{Tor86, Del87}. We note that in the case $\overline{\nu}_{0.5}(\beta)$ and $a_{0}=1.2\times10^{-10}$ m s$^{-2}$ the MOC was shown to be compact and weakly influenced by the Galactic tide, which therefore also suggests that an interpolating function that is steeper in the transition regime should be favoured.\footnote{In Newtonian dynamics, anisotropy in $b_{G}$ distribution (see Sect. \ref{classicalOC}) is introduced owing to the existence of a preferred direction, perpendicular to the Galactic midplane. By considering the MOC with $\overline{\nu}_{0.5}(\beta)$ and $a_{0}=1.2\times10^{-10}$ m s$^{-2}$ - where injection of a comet due to tides is secondary - there is also a preferred direction in the cloud, which is, although varying with time, the direction of the external field of the Galaxy. Maybe the longterm effect of the external field is to produce this kind of anisotropy in $b_{G}$ distribution for MOC comets.}

In any event, even when interpolating functions and $a_{0}$ that are in line with the Cassini data are applied, some effects of MD can be still present. Torquing of cometary perihelia owing to EFE at heliocentric distances where the Galactic tide is weak, can be important. To illustrate and quantify this effect, we plotted $\Delta L$ for the same $a=10$ kau Monte Carlo comets as in Sect. \ref{amch} but now assuming QUMOND with $\overline{\nu}_{1.5}(\beta)$ and $a_{0}=8.1\times10^{-11}$ m s$^{-2}$ instead of $\overline{\nu}_{0.5}(\beta)$ and $a_{0}=1.2\times10^{-10}$ m s$^{-2}$. One revolution of a comet now corresponds well to a Keplerian period since we are effectively in the Newtonian regime.
We have used $\alpha=1.5$, although \citetalias{Hee+15} found that $\alpha$ is constrained as $\alpha\geq2.0$, because of its numerical convenience, i.e. to speed up our numerical calculations. Our aim is to see how effective the torquing is owing to the EFE when the whole MOC is essentially in the Newtonian regime, see Fig. \ref{img:cassiniif}. In this sense $\alpha=1.5$ with $a_{0}=8.1\times10^{-11}$ m s$^{-2}$ serve us well.

\begin{figure*}\centering
\resizebox{0.65\hsize}{!}{\includegraphics{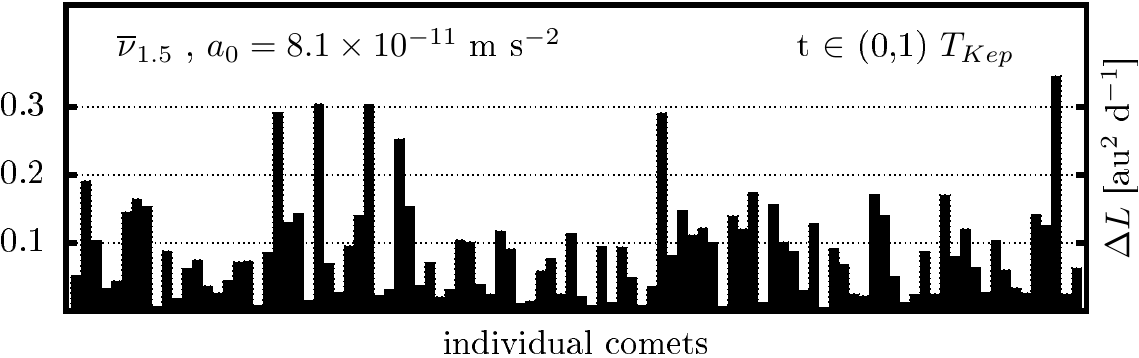}}
\caption{Histogram of $\Delta L\equiv L_{max}-L_{min}$ for 100 Monte Carlo test particles initialised at perihelia with $a=$10 kau and $q$ uniformly distributed on the interval $(15,100)$ au and then evolved in the simple model of the MOC for $T_{rev}\approx T_{Kep}\approx 1$ Myr. $\overline{\nu}_{1.5}(\beta)$ and $a_{0}=8.1\times10^{-11}$ m s$^{-2}$ were used here instead of $\overline{\nu}_{0.5}(\beta)$ and $a_{0}=1.2\times10^{-10}$ m s$^{-2}$ as in Fig. \ref{img:dl0}. A single bin corresponds to a single test particle in the simulation.
}
\label{img:dl0_nu15}
\end{figure*}

Fig. \ref{img:dl0_nu15} shows the value of $\Delta L\equiv L_{max}-L_{min}$ of the individual comets, where, again $L_{max}\equiv [L(t)]_{max}$ and $L_{min}\equiv [L(t)]_{min}$ are the maximal and the minimal value of $L(t)$ during $T_{rev}$,  assumed to be the Keplerian period $T_{Kep}$, since now $T_{rev}\approx T_{Kep}$.
We can directly compare Figs. \ref{img:dl0} and \ref{img:dl0_nu15}. On average, $\Delta L$ is naturally smaller in the case of a steeper interpolation function and smaller $a_{0}$, but extremal values in both cases are similar. This could explain why we observe comets with relatively small semi-major axes, which should be prevented from being delivered into the inner solar system due to the Jupiter-Saturn barrier \citepalias{DK11}, and at the same time, an imprint of the Galactic tide as inferred from the majority of observed comets.

\subsection{Sedna}\label{Sedna2}
We are interested in how important the torquing of perihelion is owing to EFE in the trans-Neptunian region when one substitutes $\alpha=0.5$ of $\overline{\nu}_{\alpha}(\beta)$ interpolating function and the standard value $a_{0}=1.2\times10^{-10}$ m s$^{-2}$ with larger values of $\alpha$ and $a_{0}\leq 8.1\times 10^{-11}$ m s$^{-2}$ \citepalias{Hee+15}.

\begin{figure}\centering
\resizebox{0.7\hsize}{!}{\includegraphics{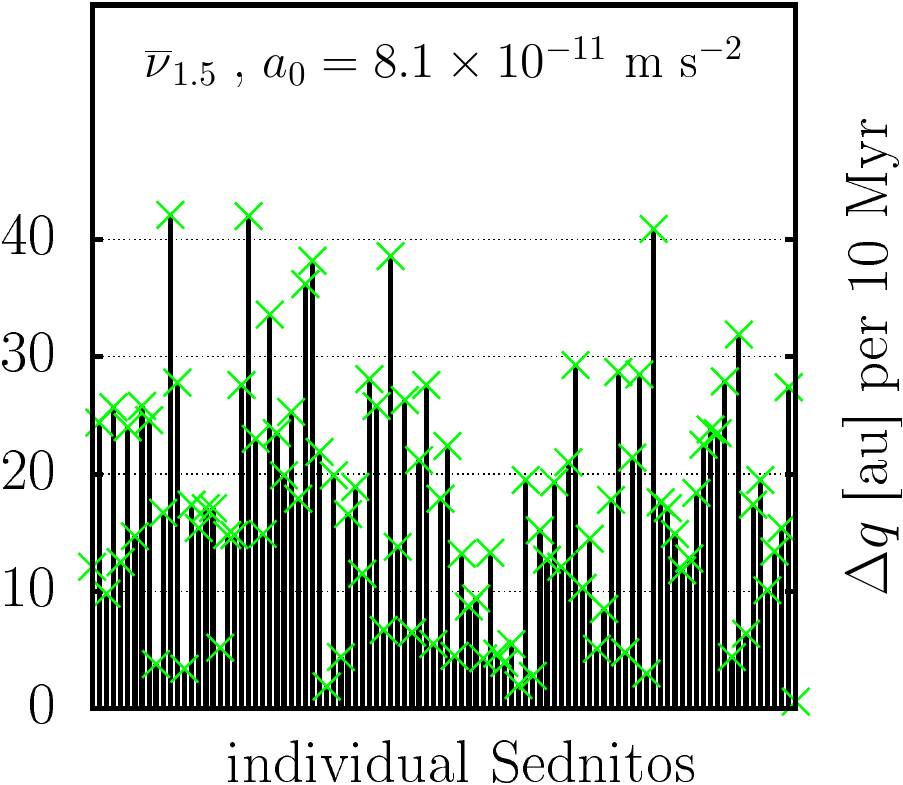}}
\caption{$\Delta q \equiv q_{max}-q_{min}$ per 10.0 Myr for 100 Sedna progenitors (Sednitos) moving under the action of EFE. $\overline{\nu}_{1.5}(\beta)$ and $a_{0}=8.1\times10^{-11}$ m s$^{-2}$ were used instead of $\overline{\nu}_{0.5}(\beta)$ and $a_{0}=1.2\times10^{-10}$ m s$^{-2}$ as in Fig. \ref{img:Sedna}. Sednitos were initialised at perihelia with $a=524$ au, uniformly distributed $q\in(25,30)$ au, $i\in(-10,10)\deg$, $\omega$, and $\Omega\in(0,2\pi)$.}
\label{img:Sedna2}
\end{figure}

We ran similar QUMOND simulation as in Sect. \ref{Sedna}, assuming $\overline{\nu}_{1.5}(\beta)$ and $a_{0}=8.1\times10^{-11}$ m s$^{-2}$.
We used the same initial orbit assignment for Sednitos as in Sect. \ref{Sedna}, except for the value of $q$ which is now a random number uniformly distributed on the interval (25, 30) au, to maximise $\Delta q$.
In Fig. \ref{img:Sedna2}, we show $\Delta q$ per 10.0 Myr for 100 simulated Sednitos. The extremal $\Delta q$ is about 50 au per 10.0 Myr. At the end of the simulation, two Sednitos had $q\sim75$ au, hence very close to the Sedna's perihelion distance. Sedna-like orbits (here simplified as specific $a$ and $q$ values) can be produced by EFE from those having a protoplanetary-disk origin in $\sim10$ Myr. 

If interpolating function $\overline{\nu}_{\alpha}$ with $\alpha\geq2.0$ would be used instead, then we can expect larger times would be necessary to produce the given $\Delta q$. We have tested this in an approximation of the EFE-induced quadrupole anomaly\footnote{The inner OC can be crudely investigated with the aid of the multipole expansion approach \citep{Mil09c,BN11}, taking into account only the dominant quadrupole term and assuming the constancy of the parameter $Q_{2}$ \citep{BN11,Hee+15}. Rotation of the external field can, in this case, be easily incorporated.} \citep{Mil09c,BN11} and the quadrupole strengths that are listed, on \cite{Hee+15}. For $\overline{\nu}_{2}$ and $a_{0}=8.1\times10^{-11}$ m s$^{-2}$ the timescale of producing given $\Delta q$ is in average, by a factor of few times greater.

\begin{figure}
\begin{center}
\resizebox{0.7\hsize}{!}{\includegraphics{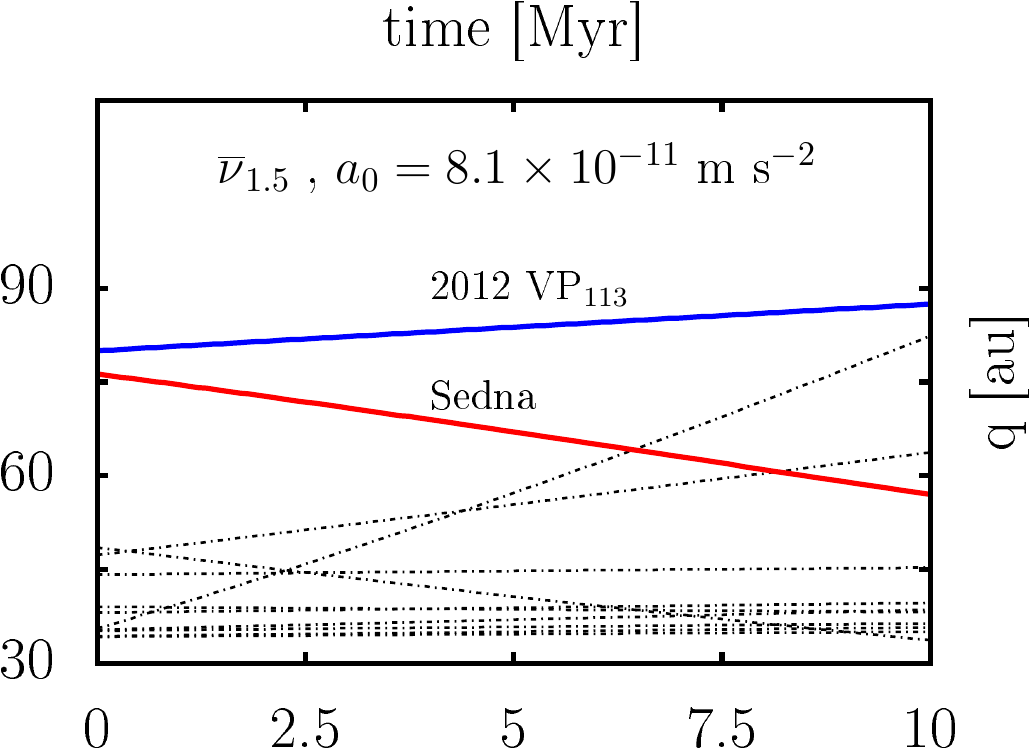}}
\caption{Perihelion distance as a function of time under action of EFE for the known population of trans-Neptunian bodies, with $a>150$ au and $q>30$ au. $\overline{\nu}_{1.5}(\beta)$, and $a_{0}=8.1\times10^{-11}$ m s$^{-2}$ were used instead of $\overline{\nu}_{0.5}(\beta)$ and $a_{0}=1.2\times10^{-10}$ m s$^{-2}$, as in Fig. \ref{img:Sednitos}. Thick solid lines are 2012 VP$_{113}$ and Sedna.}
\label{img:Sednitos2}
\end{center}
\end{figure}

In Fig. \ref{img:Sednitos2}, we depict the perihelion distance as a function of time, $q(t)$, for the known trans-Neptunian objects with $q>30$ au and $a>150$ au, in the next 10 Myr, assuming $\overline{\nu}_{1.5}(\beta)$ and $a_{0}=8.1\times10^{-11}$ m s$^{-2}$.
None of the followed objects migrates under 30 au in the next 10 Myr.

\section{Discussion and conclusion}\label{sum}
We have investigated how the (Newtonian) paradigm of a vast cometary reservoir, the Oort cloud (OC), changes in Milgromian dynamics (MD), specifically in the modified gravity QUMOND. The results are dependent on the choice of the MD interpolating function and value of the constant $a_{0}$. 

For the popular pair, $\overline{\nu}_{0.5}$ [Eq. (\ref{if3}) with $\alpha=0.5$] and $a_{0}=1.2\times10^{-10}$ m s$^{-2}$, we have found the following qualitative properties of the Milgromian OC (MOC):
\begin{itemize}
\item The observationally inferred MOC is compact with a radius $\sim15$ kau.
\item Binding energies of comets are significantly increased compared to those of the classical OC. The planetary barrier shifts significantly inward.
\item An injection of comets into the inner solar system is mainly driven by the external field effect (EFE) from the Galaxy, the specific feature of nonlinear MD, see Sect. \ref{Sec:EFE}. The Galactic tide can be, owing to small heliocentric distances of MOC bodies, neglected.
\item EFE-induced injection of comets is very efficient and the cometary influx can be significantly larger than in the Newtonian case, if we assume the zeroth approximation of the same, Newtonian, source population and its distribution in both frameworks. 
\item The orbit of a body with a proto-planetary disk origin can, under the action of EFE, be transformed into a Sedna-like orbit on a timescale of several Myr.
\end{itemize}
Trans-Neptunian bodies in Sedna-like orbits are not ``fossil'' objects with frozen perihelia (like in the Sun-in-a-cluster model) but rather they could repeatedly migrate through the inner solar system as EFE raises and lowers their perihelia repeatedly.

During the preparation of this paper, \citetalias{Hee+15} published constraints on various commonly-used MD interpolating function families. The constraints are based on the Cassini spacecraft radio tracking data \citep{Hee+14}. Many popular MD interpolating functions have been proven incompatible with the data, including $\overline{\nu}_{0.5}$.
Adopting $\overline{\nu}_{\alpha}(\beta)$ with $\alpha\gtrsim1.5$ and $a_{0}\leq 8.11\times10^{-11}$ m $s^{-2}$ in line with \citetalias{Hee+15}, the MD-regime is essentially suppressed at any distance from the Sun owing to EFE, see Fig. \ref{img:cassiniif}. The cloud is, in this case, very much Newtonian in its overall size, binding energies of comets and operation of the Jupiter-Saturn barrier. However, even in this case, EFE substantially torques orbits in the inner parts of the cloud where the tidal force is weak, with the potential to transform primordial orbits to Sedna-like orbits, as was shown in the case $\alpha=1.5$ and $a_{0}=8.11\times10^{-11}$ m $s^{-2}$ on the timescale of $\sim10$ Myr. Steeper interpolating functions imply larger timescales for these transformations.

To sum up, if the results presented in \citet{Hee+14} are correct, and there is no other hidden dynamical effect acting on the spacecraft, then the results presented in sections \ref{MilgromianOC} - \ref{XXZ} are only of academic character. Still, it is instructive to see how the MOC changes with a varying description of the transition regime.

We further discuss the MOC in line of the new constraints on the MD interpolating function.
We emphasise that the influence of EFE on inner OC bodies and Centaurs, Kuiper belt objects, and scattered disk objects in high$-a$ orbits is even under these circumstances substantial. Consequently Sedna-like orbits and orbits of large semi-major axis Centaurs are easily comprehensible in MD. In MD, they both belong to the same population, just in different modes of their evolution.

MD could eventually shed light on many open problems in the cis and trans-Neptunian region. Besides the already mentioned puzzling orbits of Sedna and 2012 VP$_{113}$, the clustering in argument of perihelion, $\omega$, near $\omega\approx 0\deg$, for bodies in orbits with $q>30$ au and $a>150$ au \citep{TS14}, and the origin of high-$a$-Centaurs \citep{Gom+15}, could be elucidated in MD. 
With regards to $\omega$-clustering, EFE would manifest in this region through an anomalous force that increases with heliocentric distance and is aligned with the direction to the Galactic centre \citep{Mil09c,BN11}. Hence bodies, that are protected from encountering the planets frequently, should bear an imprint of EFE, similarly, as if there was a distant massive body hidden deep in the OC. In MD, one could expect nodal ($\Omega$), or eventually both nodal and apsidal ($\omega+\Omega$), confinement (Pau\v{c}o 2016, in preparation). The fact that a subsample of the stable objects (with $a>250$ au) is actually clustered in the physical space was recently shown in \cite{BB16}. 

EFE is an important dynamical agent, raising and lowering perihelia in the inner parts of the outer solar system very effectively, with no such counterpart in Newtonian dynamics. Thus, we could intuitively expect MOC, and especially its inner part, to be more populous at the formation phase than the classical OC, as planetesimals with mildly pumped semi-major axes ($a\sim$ 0.1 - 1 kau) could have their perihelia lifted sufficiently rapidly to be protected from being ejected or captured by planets. Also, we could expect this primordial outward migration to be followed by a period of high influx of interplanetary material, after which (or after several such cycles) this inner region was radically depleted. Here timing is important because this phenomenon could coincide with the late heavy bombardment, hinted at by the Moon's petrology record \citep{Har+00}, at $\sim$700 Myr after planets formed. Although this kind of event is rather abrupt and of relatively short duration, it was well accounted for in the Newtonian framework with the model of rapid migration of the outer planets \citep{Gom+05}. We plan to investigate this topic in a subsequent work.

It is questionable whether the primordial disk mass and OC-to-scattered-disk population ratio problems arise in the context of MD since nobody has ever simulated solar system formation and evolution (with its outermost parts) in MD. EFE torquing is important in the context of the (re)distribution of material within the cloud, which could be then expected to be different in MD to that in Newtonian dynamics. The preference for high semi-major axis orbits (where tides are sufficiently strong) in the classical OC does not need to be so eminent in MOC. In the perihelion distance, $q$, vs. semi-major axis, $a$, diagram, where in the classical OC theory there is more or less empty space at $q\gtrsim 100$ au, $a\sim1000$ au, we expect some residual population in MD. In the future, this could be tested against observations. Also, a simulation similar to that in sections \ref{Sedna} and \ref{Sedna2}, but including the outer planets and more Sednitos, would yield steady populations of bodies with $q>30$ au and  $q<30$ au (high-$a$ Centaurs), after some time, which could be tested against observations on a similar basis to \cite{Gom+15}. There is obviously some tension between the theory and observations in the Newtonian framework \citep{Gom+15}.

At this stage, we cannot claim MD to be self-consistent solution of the puzzles that bother classical OC theory, but it has been shown that it can well form a new, testable, paradigm with a specific signature in the outer parts of the solar system.

\begin{acknowledgements}
We are thankful to Leonard Korno\v{s} and Lubo\v{s} Neslu\v{s}an for valuable discussions on orbital integrators and the classical Oort cloud. We also thank the referee, Rodney Gomes, for an open-minded review and comments, which helped to improve the clarity of the paper and also inspired us to realize an additional motivation for consideration of Milgromian dynamics in the solar system.
J.K. is supported by the Slovak National Grant Agency VEGA, grant No. 1/067/13.
\end{acknowledgements}

\begin{table*}
\caption{Original barycentric orbital elements of Sect. \ref{simul} near-parabolic comets. These 31 comets were identified as dynamically new (assuming Newtonian dynamics) in the sample of \citetalias{DK11}. Presented orbital elements are expected values retrieved from \cite{Kro14}, errors are omitted. Successive columns are: comet designation, osculation date, perihelion distance, eccentricity, inclination, argument of perihelion, longitude of ascending node (all angles in equinox J2000.0), semi-major axis and perihelion passage time.}\label{comets}
\begin{center}
\begin{tabular}{lcccrrrrc}
\hline\hline
Comet & Epoch & $q$\tablefootmark{$\dag$} & $e$ & $i$\tablefootmark{$\dag$} & $\omega$\tablefootmark{$\dag$} & $\Omega$\tablefootmark{$\dag$} & $a$\tablefootmark{$\dag$} & $T$\tablefootmark{$\ddag$}\\
\hline
[...] & [yyyymmdd] & [au] & [...] & [$\deg$] & [$\deg$] & [$\deg$] & [kau] & [yyyymmdd]\\
\hline
C/1974 V1 & 16670721 & 6.02 & 0.99989464 & 60.9 & 151.8 & 226.1  & 57.110 & 19740808\\
C/1978 A1	&	16701212	&	5.61	&	0.99978957	&	116.9	&	343.4	&	211.7	&	26.652	& 19770722\\
C/1978 G2	&	16710809	&	6.28	&	1.00014083	&	153.2	&	229.7	&	72.2	&	-44.603	& 19780826\\
C/1984 W2\tablefootmark{$\diamond$}	&	16820212	&	4.00	&	0.99991890	&	89.3	&	255.3	&	250.2	&	49.383	& 19850929\\
C/1987 W3	&	16850706	&	3.32	&	0.99991866	&	76.8 	&	195.1	&	198.4	&	40.850	& 19880119\\
C/1988 B1	&	16811015	&	5.03	&	0.99989942	&	80.6	&	124.2	&	325.2	&	50.025 & 19870319\\
C/1992 J1	&	16910824	&	3.00	&	0.99991839	&	124.3	&	83.5	&	203.3	&	36.765	& 19930904\\
C/1997 A1	&	16950405	&	3.16	&	0.99993098	&	145.1	&	40.0	&	135.7	&	45.830	& 19970620\\
C/1997 BA6\tablefootmark{$\diamond$} &	16970213	&	3.44	&	0.99989050	&	72.6	&	285.9	&	317.7	&	31.417	& 19991128\\
C/1999 J2	&	16910317	&	7.11	&	0.99984298	&	86.4	&	127.1	&	50.1	&	45.310 & 20000405	\\
C/1999 K5	&	16980208	 &	3.25	&	0.99993034	&	89.5	&	241.5	&	106.3	&	46.707	& 20000703\\
C/1999 U4	&	16960618	&	4.89	&	0.99984462	&	52.1&	77.8	&	32.4	&	31.447 & 20011029\\
C/2000 A1	&	16861029	&	9.74	&	0.99960423	&	24.6&	14.3	&	111.9	&	24.612 & 20000715	\\
C/2001 C1	&	16960906	&	5.11	&	0.99991858	&	68.9	&	220.0	&	33.8	&	62.775 & 20020328	\\
C/2001 K3	&	16990214	&	3.07	&	0.99990440	&	52.0	&	3.4&	289.8	&	32.103 & 20010423\\ 
C/2001 K5	&	16970325	&	5.19	&	0.99994997	&	72.5	&	47.1&	237.5	&	103.734 & 20021011	\\
C/2002 A3	&	16960906	&	5.14	&	0.99989342	&	48.1	&	329.6	&	136.7	&	48.263	& 20020425\\
C/2002 J4	&	17000817	&	3.64	&	0.99987674	&	46.5	&	230.7	&	70.9	&	29.516 & 20031003	\\
C/2002 L9	&	16950224	&	7.04	&	0.99974285	&	68.4	&	231.4	&	110.5	&	27.360	& 20040405\\
C/2003 G1	&	16971120	&	4.92	&	0.99993277	&	66.8	&	11.5	&	246.1	&	73.206 & 20030204	\\
C/2003 S3	&	16920530	&	8.13	&	0.99971781	&	151.5	&	154.4	&	226.3	&	28.802	& 20030409\\
C/2004 P1	&	16960509	&	6.02	&	0.99981290	&	28.8	&	16.5	&	284.2	&	32.185 & 20030809	\\
C/2004 T3	&	16901227	&	8.87	&	0.99959502	&	71.9	&	259.7	&	50.4	&	21.891 & 20030414\\
C/2004 X3	&	17010414	&	4.39	&	0.99994104	&	81.2	&	202.4	&	343.0	&	74.460	& 20050618\\
C/2005 B1\tablefootmark{$\diamond$} &	17040109	&	3.21	&	0.99998720	&	92.5	&	103.1&	195.6	&	250.627	& 20060222\\
C/2005 G1	&	17001215	&	4.95	&	0.99991785	&	108.4	&	113.9	&	299.6	&	60.314 & 20060226\\
C/2005 K1\tablefootmark{$\diamond$} &	17021205	&	3.69	&	0.99996944	&	77.8	&	135.0	&	106.3	&	120.773	& 20051121\\
C/2005 Q1	&	16971120	&	6.40	&	0.99985473	&	105.3	&	44.8	&	87.7&	44.053 & 20050826	\\
C/2006 E1	&	16991001	&	6.04	&	0.99980395	&	83.2	&	232.8	&	95.1	&	30.788 & 20070106	\\
C/2006 K1	&	17030404	&	4.42	&	0.99992817	&	53.9	&	296.5	&	72.2&	61.576 & 20070720	\\
C/2007 Y1	&	17050722	&	3.34	&	0.99988596	&	110.1	&	357.1	&	133.1	&	29.317	& 20080318\\
\end{tabular}
\end{center}
\tablefoot{
\tablefoottext{$\diamond$}{Non-gravitational effects are accounted in the orbit determination, see \citetalias{DK11}.}
\tablefoottext{$\dag$}{These orbital elements rounded off
from those of \cite{Kro14}.}
\tablefoottext{$\ddag$}{.dddddd part is omitted compared to \cite{Kro14}.}
}
\end{table*}

%
%

\bibliographystyle{aa}
\bibliography{aa}

\end{document}